\documentclass[aps,prx,reprint,longbibliography]{revtex4-1}%
\usepackage{epsfig}
\usepackage{latexsym,pslatex, times}

\usepackage{bm, float, lipsum}
\usepackage{amsmath}
\usepackage{amsfonts}
\usepackage{amssymb}
\usepackage{mathrsfs, bbold}
\usepackage{color}
\usepackage{graphicx}%

\usepackage[normalem]{ulem}

\providecommand{\U}[1]{\protect\rule{.1in}{.1in}}
\begin{document}
\author{N. J. Harmon}
\email{nh140@evansville.edu} 
\affiliation{Department of Physics, University of Evansville, Evansville, Indiana
47722, USA\\Department of Physics and Astronomy and Optical Science and Technology Center, University of Iowa, Iowa City, Iowa
52242, USA}
\author{M. E. Flatt\'e}
\email{michael\_flatte@mailaps.org}
\affiliation{Department of Physics and Astronomy and Optical Science and Technology Center, University of Iowa, Iowa City, Iowa
52242, USA\\
Pritzker School of Molecular Engineering, University of Chicago, Chicago, Illinois, 60637, USA\\Department of Applied Physics, Eindhoven University of Technology, P.O. Box 513, 5600 MB, Eindhoven, The Netherlands}
\date{\today}
\title{Theory of Single Photon Detection by a Photoreceptive Molecule and a Quantum Coherent Spin Center}
\begin{abstract}
The long spin coherence times in ambient conditions of color centers in solids, such as nitrogen-vacancy (NV$^{-}$) centers in diamond, make these systems attractive candidates for quantum sensing. Quantum sensing provides remarkable sensitivity at room temperature to very small external perturbations, including magnetic fields, electric fields, and temperature changes. A photoreceptive molecule, such as those involved in  vision, changes its charge state or conformation in response to the absorption of a single photon. We show the resulting change in local electric field modifies the properties of a nearby quantum coherent spin center in a detectable fashion.  Using the formalism of positive operator values measurements (POVMs), we analyze the photo-excited electric dipole field and, by extension, the arrival of a  photon based on a measured readout, using a fluorescence cycle, from the spin center. We determine the jitter time of photon arrival and the probability of measurement errors.  We predict that configuring multiple independent spin sensors around the photoreceptive molecule would dramatically suppresses the measurement error.
\end{abstract}
\maketitle

\section{Introduction} 
Single-photon detectors require high-efficiency absorption of individual photons in a material combined with a dramatic amplification of the resulting material perturbation \cite{OConnor1984,Hadfield2009}. For solid-state single-photon detectors the usual initial perturbation is a change in conductivity, through the conversion of the photon into an electron-hole pair, or a change in temperature, through the conversion of the photon energy into heat. In semiconductor avalanche diodes \cite{McIntyre1966,Stillman1977} the initial conductivity change is amplified through a cascade wherein an initial electronic carrier is accelerated to sufficiently high energies to relax to lower energies while simultaneously generating another electron-hole pair. Under proper conditions this impact ionization leads to a macroscopically-measurable conductivity. For superconducting nanowire detectors the macroscopic change is the conversion of a region of a thin superconducting wire from the superconducting state to the normal state, producing a large resistance change in the device \cite{Eisaman2011,Natarajan2012}. For both technologies the amplification process proceeds through rapid, incoherent electronic events designed to produce an irreversible event corresponding to a photodetector `click'. The use of quantum coherent sensors promises considerably greater sensitivity and control of the single photon detection process, as the evolution of one quantum coherent  sensor may or may not interact with the evolution of another quantum coherent sensor, and both can be independently influenced by the response of a photoreceptive object. 

Here we analyze a particular case of quantum coherent single-photon detection, in which a quantum-coherent spin center detects the response of a photoreceptive molecule, which changes its conformation upon absorbing a single photon thus producing a static modification of the electric dipole moment of the molecule. Electric-field detection at room temperature with a single nitrogen-vacancy (NV$^-$) quantum coherent spin center has been demonstrated  \cite{Dolde2011}, so for specificity we use parameters associated with the NV$^-$ center, although other spin centers are known to be more sensitive to electric fields due to larger spin-orbit interactions. For example, the Mn in GaAs spin center is predicted to be $10^4$ times as sensitive to electric fields as the NV~center\cite{Tang2006}. 
Our proposed approach relies on the construction of positive operator valued measurement (POVM) operators \cite{BarnettBook}  to discriminate the discrete on/off states of an electric field with minimal error. The discrimination error can be minimized by measuring the NV state at an optimal time determined theoretically \cite{Chaudhry2015}. 
We predict that the POVM approach provides a far faster detection protocol than reported in Ref.~\onlinecite{Dolde2011}. Furthermore the introduction of multiple quantum spin centers provides the ability to measure the jitter in the single-photon detection event and to reduce the measurement error (false positive and false negative), with an exponential dependence on the number of quantum spin centers. These techniques, while applied here to the response of the photoreceptor molecule, should also have broad application for the detection of other sources of localized static electric fields, such as charges in single-electron transistors \cite{Kastner1992}.

The theory of quantum sensing using a photoreceptive molecule and an NV$^-$ spin is detailed in Section \ref{section:theory}. The essence of the theoretical approach relies on the theory of minimal error non-orthogonal state discrimination. We note that if unambiguous detection of the single photon's presence or absence is required, for example for quantum cryptography purposes, then instead one would use unambiguous state discrimination theory based on POVMs \cite{BarnettBook}. However, the minimal error for that formalism is guaranteed to be larger than the minimal error for non-orthogonal state discrimination, so absent this requirement a superior approach is to perform minimal-error non-orthogonal state discrimination \cite{Herzog2004}.
Section \ref{sec:casestudies} examines three limits for which analytic solutions for the spin evolution and resulting error rates are obtained. 
Section \ref{section:optimization} explores how measurements of the NV$^{-}$ state can be optimized to reduce error probability. The error probability is periodic in time and reaches a minimal value at the time $t_{min}$. If the decoherence rate is slower than the spin precession rate in the photo-induced electric dipole field then $t_{min}$ can be expressed as an analytic function of the system parameters. 
Our analysis suggests a specific experimental pulse sequence which is effective in measuring the NV$^{-}$ state, and hence the probability of single photon detection. 
An additional and dramatic improvement in the error rate can be achieved by considering multiple independent NV$^{-}$ spin sensors.
Section \ref{section:tradeoffs} discusses various figures of merit, what trade-offs take place when optimizing one figure of merit over another, and how the arrival time of a photon can be determined. These theoretical calculations provide a framework for  single-photon detection using quantum coherent spin sensors of local electric fields. 


\begin{figure}[ptbh]
 \begin{centering}
        \includegraphics[width=\columnwidth]{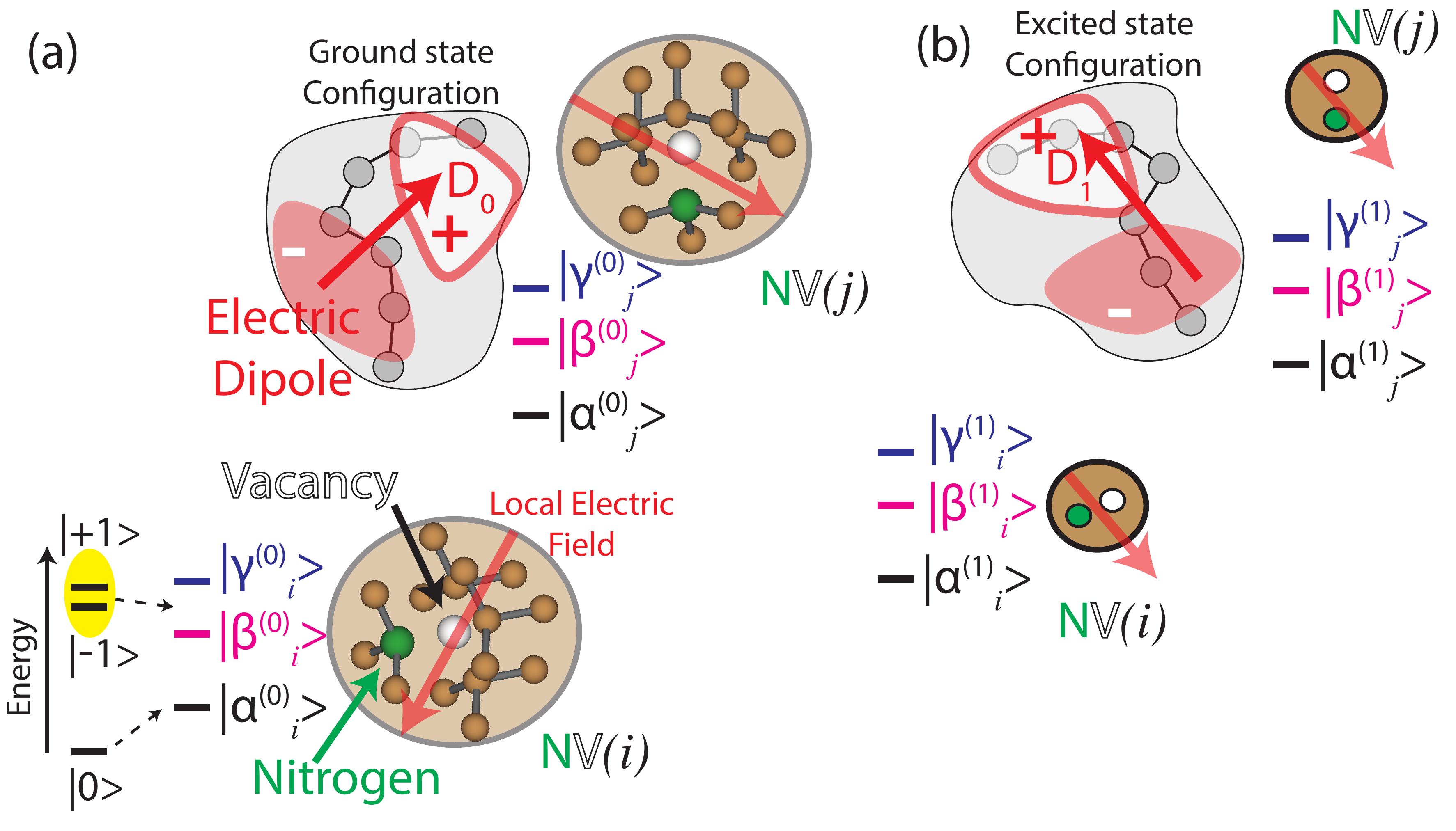}
        \caption[]
{Toy model of single photon detection using a photoreceptor and a nitrogen vacancy (NV$^-$) center in diamond. Also shown are the ground state (triplet) energy levels for the NV spin with a magnetic field in the axial direction ($\hat{z}$, pointing from N to V)  and including zero-field splitting. The separations of energy levels are not drawn to scale. (a) No photon is absorbed so the spin evolves according to the photoreceptor's electric dipole field in the photoreceptor's ground state. (b) Photon is absorbed which induces an altered electric dipole field (in magnitude and direction) and modifies the coherent precession of the NV$^-$ spin within the three ground-state sublevels. The two configurations are, in general, non-orthogonal, suggesting that POVM's provide the optimal framework for measurement.}\label{fig:NVsensor} 
        \end{centering}
\end{figure}

\section{Theory of Detection Error Probability}\label{section:theory}

The proposed model of single photon detection can be visually understood by examining Fig.~\ref{fig:NVsensor}. 
In Fig.~\ref{fig:NVsensor}(a) the NV spin is initiated in one of the three ground state sublevels $|0\rangle$, $|+1\rangle$, and $|-1\rangle$, where the number indicates the axial projection of the spin (quantum number $m$). In the presence of external electric and magnetic fields these three sublevels are mixed, and the energies modified, into three new levels $|\alpha\rangle$, $|\beta\rangle$, and $|\gamma\rangle$, which depend on the direction and values of the local electric and magnetic fields. If the local electric field changes, then the wavefunctions of these eigenstates and their energies also change, and so they are labeled with superscripts $(0)$ and $(1)$ in Fig.~\ref{fig:NVsensor}. Some element of this change in eigenstates and energies, in response to an altered electric field representing the absorption of a photon by the photoreceptive molecule,  must be measured to determine if a photon was absorbed. To perform such a measurement the spin should be initialized in a known coherent state, and its subsequent evolution, controlled by these eigenstates and energies, determined sufficiently to distinguish the two possible local electric fields.

The Hamiltonian, which will be introduced in Eq.~(\ref{hamiltonian}), describes how $|\alpha\rangle$, $|\beta\rangle$, and $|\gamma\rangle$ change in response to applied electric and magnetic fields. The electric field couples the two states $|+1\rangle$, and $|-1\rangle$, and leaves the $|0\rangle$ state uncoupled to the others, so we assume preparation in some linear combination of $|+1\rangle$, and $|-1\rangle$. This can be achieved through optical polarization of the spin through a fluorescence cycle, followed by a coherent manipulation to place the spin in the desired initial state. 
In Fig.~\ref{fig:NVsensor}(a) no photon has been absorbed. Given the electric dipole moment from the photoreceptive molecule in its unexcited state, the spin evolves and is represented by the ket $a_0(t) | +1 \rangle + b_0(t) | -1 \rangle$ at a later time. Alternatively in Fig.~\ref{fig:NVsensor}(b) a photon has been absorbed by the photoreceptive molecule which gives rise to a different electric dipole moment and field. A spin initiated in the same manner as Fig.~\ref{fig:NVsensor}(a)  will evolve according to this altered spin Hamiltonian and at some time later be in the state $a_1(t) | +1 \rangle + b_1(t) | -1 \rangle$. 
By performing optical measurements the NV spin state would be ascertained, although with a certain amount of uncertainty because the kets associated with the two final states will not be orthogonal in general. 
The spin evolution under either of the circumstances shown in Figure \ref{fig:NVsensor} (either absorbing or not absorbing a photon) and a determination of which measurements should take place, and when, in order to minimize the error probability, are developed in this section.

Consider the ground state spin Hamiltonian, $\mathscr{H}_{gs}$, of a diamond NV-center triplet spin ($S = 1$) where there is zero-field splitting ($2\pi \hbar D_{gs}$), an electric dipole term from both strain ($\bm{\sigma}$) and the external electric field ($\bm{E}$) where dipole moments, $d_{gs, k}$, are induced by the spin-orbit interaction, and a Zeeman interaction from an axial magnetic field. 
Note that the strain field can be expressed as a vector since diamond's inversion symmetry reduces the number of deformation constants in the strain tensor from six to three \cite{Yu3ed}.
Strain is ignored here but the following results apply equally well to strain detection.
The ground state Hamiltonian is \cite{Doherty2013, Ajisaka2016}
\begin{widetext}
\begin{equation}
\mathscr{H}_{gs}   + \mathscr{H}_z = (2 \pi \hbar D_{gs} + d_{gs, ||} E_z) S_z^2
 + d_{gs, \perp} \Big[E_x (S_x^2 - S_y^2)+ E_y (S_x S_y + S_y S_x)  \Big]  
 + g \mu_B B_z S_z,\label{hamiltonian}
\end{equation}
or, in matrix form,
\begin{equation}
\mathscr{H}_{gs} + \mathscr{H}_z = \left(
\begin{array}{ccc}
\mathcal{D} + \mathcal{B}_z & 0 & \mathcal{E}_{\perp}^*  \\
 0 & 0 & 0 \\
\mathcal{E}_{\perp} & 0 & \mathcal{D} -\mathcal{B}_z \\
\end{array}
\right).
\end{equation}
with $\mathcal{D}  =2 \pi \hbar  D_{gs}  + d_{gs, ||} E_z $,  $\mathcal{E}_{\perp} =  d_{gs, \perp} (E_x + i E_y)$, and $\mathcal{B}_z = g \mu_B B_z$.
The energy eigenvalues of the ground state Hamiltonian are
\begin{equation}
\varepsilon_0 = 0, \quad \varepsilon_{\pm 1} = \mathcal{D}
\pm  \Delta \varepsilon , \qquad  \Delta \varepsilon = \sqrt{ |\mathcal{E}_{\perp}|^2+ \mathcal{B}_z^2 }.
\end{equation}
\end{widetext}
An abbreviated energy diagram of the NV center is depicted in Figure 1.

As long as there is not a non-axial magnetic field (that is $B_x$ and $B_y$ vanish) then the $m_s =0$ state does not couple to either $m_s = \pm 1$ state.
We consider dynamics only between the $m_s = \pm 1$ states which allows us to reduce the dimension of $\mathscr{H}_{gs}$ from three to two:
\begin{equation}
\mathscr{H}_{gs} + \mathscr{H}_z = \left(
\begin{array}{cc}
\mathcal{D} + \mathcal{B}_z  & \mathcal{E}_{\perp}^*  \\
\mathcal{E}_{\perp}  & \mathcal{D} - \mathcal{B}_z\\
\end{array}
\right).
\end{equation}
Note that the non-axial electric fields couple the $m_s = \pm 1$ states ($\Delta m_s  = 2$) much like a magnetic field couples $m_s = \pm\frac{1}{2}$ states ($\Delta m_s  = 1$) for $s = \frac{1}{2}$. A non-axial magnetic field would also couple $m_s = 0 $ to the $m_s = \pm 1$ states. 
We assume $\mathcal{E}_{\perp}$, when present, is due to an externally applied electric field for which we know the magnitude and direction.

\begin{figure}[ptbh]
 \begin{centering}
        \includegraphics[width=\columnwidth]{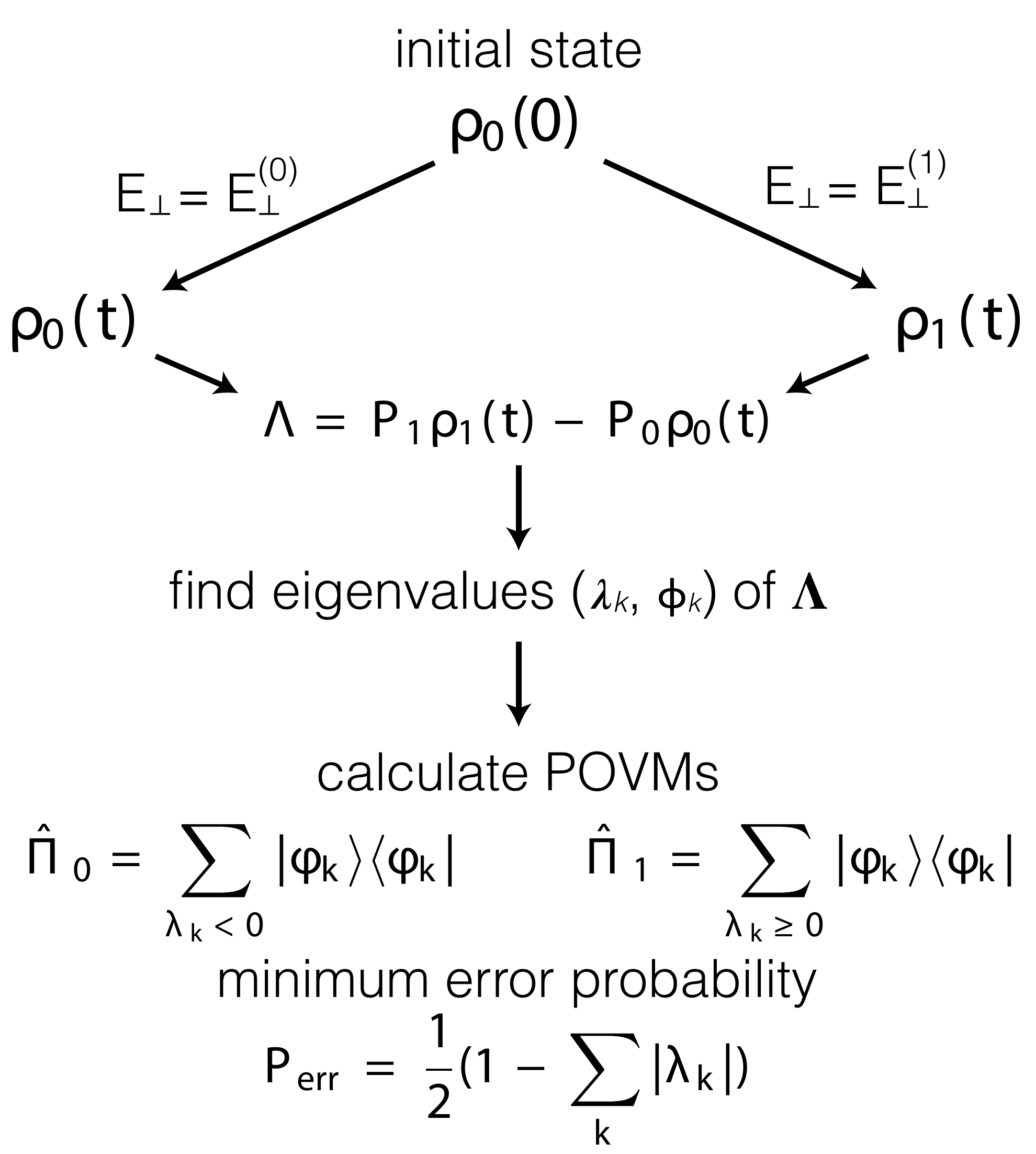}
        \caption[]
{Flow chart for the calculation of the minimum error probability $P_{\text{err}} $.}\label{fig:flowchart} 
        \end{centering}
\end{figure}

The dynamics of the NV spin are determined by solving the Liouville-von Neumann equations:
\begin{equation}\label{eq:SLE}
\frac{\partial \rho}{\partial t} = -\frac{i}{\hbar} [\mathscr{H}_{gs} + \mathscr{H}_z, \rho] +    \hat{L}_{d} \rho  \hat{L}_{d} ^{\dagger}  - \frac{1}{2} \{ \hat{L}_{d} ^{\dagger}  \hat{L}_{d} , \rho \} ,
\end{equation}
where the first term on the right hand side describes the coherent evolution of the spin and the other terms generate decoherence processes through the Lindblad operator $\hat{L}_d$. The timescale of decoherence is $1/\kappa$.
The calculation process described in the remainder of this section is summarized in the flowchart of Figure \ref{fig:flowchart}.

The theory of quantum-state discrimination \cite{Helstrom1976, Holevo1982} .                                                                                                                                                                                                                                                                                                             informs us which measurement operators minimize the error in choosing which state the defect spin occupies given a measurement value. We use a Hermitian operator,
\begin{equation}
\hat{\bm{\Lambda}} = P_1 \rho_1(t) - P_0 \rho_0(t),
\end{equation}
where $P_1$ and $P_0$ are the \emph{a priori} probabilities of there the electric field being $\mathcal{E} = \mathcal{E}_0$ or $\mathcal{E} = \mathcal{E}_1 = \mathcal{E}_0 + \Delta\mathcal{E}$, respectively. 
$\rho_0$ and $\rho_1$ correspond to the time-dependent density matrices anticipated for those respective fields.
We then seek to construct a \emph{positive operator valued measure} (POVM).
Since we are dealing with two states, the optimal POVM is also a projective measure \cite{Bergou2015, Chaudhry2015}.
Appendix A provides a brief review of POVMs. 
The POVM operators are, when the eigenvalues of $\hat{\bm{\Lambda}}$ are $\lambda_k$,
\begin{equation}\label{eq:POVM}
\hat{\bm{\Pi}}_ 0 = \sum_{\lambda_k < 0} | \phi_k \rangle \langle \phi_k |, \qquad 
\hat{\bm{\Pi}}_ 1 = \sum_{\lambda_k \ge 0} | \phi_k \rangle \langle \phi_k |,
\end{equation}
where $|\phi_k\rangle$ are eigenstates of $\hat{\bm{\Lambda}}$.  
Measurement in the system is carried out by either $\hat{\bm{\Pi}}_0$ or $\hat{\bm{\Pi}}_1$ `clicking'. For our purposes a click of $\hat{\bm{\Pi}}_0$ signifies no electric field and no chromophore absorption and a click of $\hat{\bm{\Pi}}_1$ signifies the opposite.
The minimum error probability is given by the sum of false positive and false negatives as
\begin{equation}
P_{\text{err}} = P_0 \text{Tr}(\rho_0 \hat{\Pi}_1)  +  P_1 \text{Tr}(\rho_1 \hat{\Pi}_0)
\end{equation}
but can also be shown to be expressed simply in terms of $\hat{\bm{\Lambda}}$ eigenvalues as \cite{Bergou2010}
\begin{equation}
P_{\text{err}}  = \frac{1}{2} (1 - \sum_k |\lambda_k |).
\end{equation}

In the next section we determine $P_{\text{err}} $ for three scenarios where analytic results are particularly simple and elucidating: (1) electric field detection neglecting decoherence, (2) electric field detection including decoherence, and (3) electric field detection with an axial magnetic field. For the first two cases the magnetic field is negligible, and when decoherence is included it is the fluctuations of the transverse electric field are that are the source of decoherence. 

The strength of the electric field generated by the photoreceptive molecule depends on the change in the electric dipole for the molecule and the distance between the photoreceptive molecule and the NV centers. For one characterized photoreceptor molecule, disperse red 1, the configuration is a donor and acceptor group that are separated by an azobenzene group. Under illumination the azobenzene group changes from the trans to the cis configuration, reorienting the positions of the donor and acceptor groups and changing their separation\cite{Atassi1998,Barrett2007}. For this molecule the electric dipole therefore changes upon illumination in both amplitude (from 9~Debye to 6~Debye) and in direction. This effect has been observed through the gate doping of a graphene transistor\cite{Kim2012}, which is well described theoretically\cite{Shashikala2012}. The use of graphene field effect transistors to read out a photodetector based on these photoreceptive molecules has also been proposed\cite{Leonard2017}. The electric fields associated with these changes in electric dipole correspond to electric fields of  $\sim 10^6$~V/m for photoreceptor-NV separations $\sim 10$~nm, and electric fields $\sim 10^7$~V/m for photoreceptor-NV separations $\sim 5$~nm. Photoreceptive molecules with even larger electric dipoles should be feasible to synthesize. 
 
\section{Three Case Studies}\label{sec:casestudies}

Determining the error probability, $P_{err}$, as mapped out in the last section is performed in general by numerically solving Eq.~(\ref{eq:SLE}) and then following the steps outlined in Fig.~\ref{fig:flowchart}.
In this section, three cases are examined where analytic results are obtained: 
(1) negligible decoherence and negligible magnetic field
(2) negligible decoherence
(3) negligible magnetic field.

For reasons discussed in Appendices \ref{appendixC} and \ref{appendixD}, when including the magnetic field and decoherence the solutions `entangle' the $\mathcal{B}_z$ and $\kappa$ and no simple expression for $\rho$ can be written. 
In these cases, we opt to numerically solve Eq. \ref{eq:SLE}. For these calculations the parameters in Table~\ref{NVvalues} are used. 
\begin{table}[h]
\centering
\begin{tabular}{ccc}
\hline
\hline
Parameter   &  Value & Reference \\
\hline
$D_{gs}$ & 2.87 GHz & \cite{Loubser1977, Loubser1978}   \\
$d_{gs, ||}$ & 2 $\pi \hbar \times$ 0.0035  Hz m/V  & \cite{vanOort1990}  \\
$d_{gs, \perp}$ & 2 $\pi \hbar \times$ 0.17  Hz m/V & \cite{vanOort1990}  \\
$T_2 = 1/\kappa$ & 10 $\mu$s   & \cite{Balasubramanian2009, Rosskopf2014} \\
$T_1$ &  $\infty$ &  \\
$P_0$ &  $1/2$ &  \\
$P_1$ &  $1/2$ &  \\
\hline
\hline
\end{tabular}
\caption{\label{NVvalues}Ground state energy parameters and spin relaxation times, along with references for experimental measurements of these values.}
\end{table}

To proceed, the spin initialization, $\rho(0) = \rho_0(0) = \rho_1(0)$ must be specified.
For reasons to be addressed later we prepare, at $t = 0$, a particular quantum state, $|+1\rangle$,  described by the density matrix
\begin{eqnarray}\label{eq:init}
  \rho_0(0) = 
  \left( {\begin{array}{cc}
   1 & 0  \\
   0 & 0 \\
  \end{array} } \right).
\end{eqnarray}

\subsection{Calculation of Minimum Error Probability: No Decoherence \& $\mathcal{B}_z = 0$}\label{sec:case1}

In this simplest of cases Eq. \ref{eq:SLE} is readily solved using
\[
\rho(t) =  U(t)
\rho(0)  U^{\dagger}(t),
\]
where $U(t)$ is the unitary evolution operator $\text{exp} (-i \mathscr{H}_{gs} t/\hbar)$.
\begin{figure}[ptbh]
 \begin{centering}
        \includegraphics[width=\columnwidth]{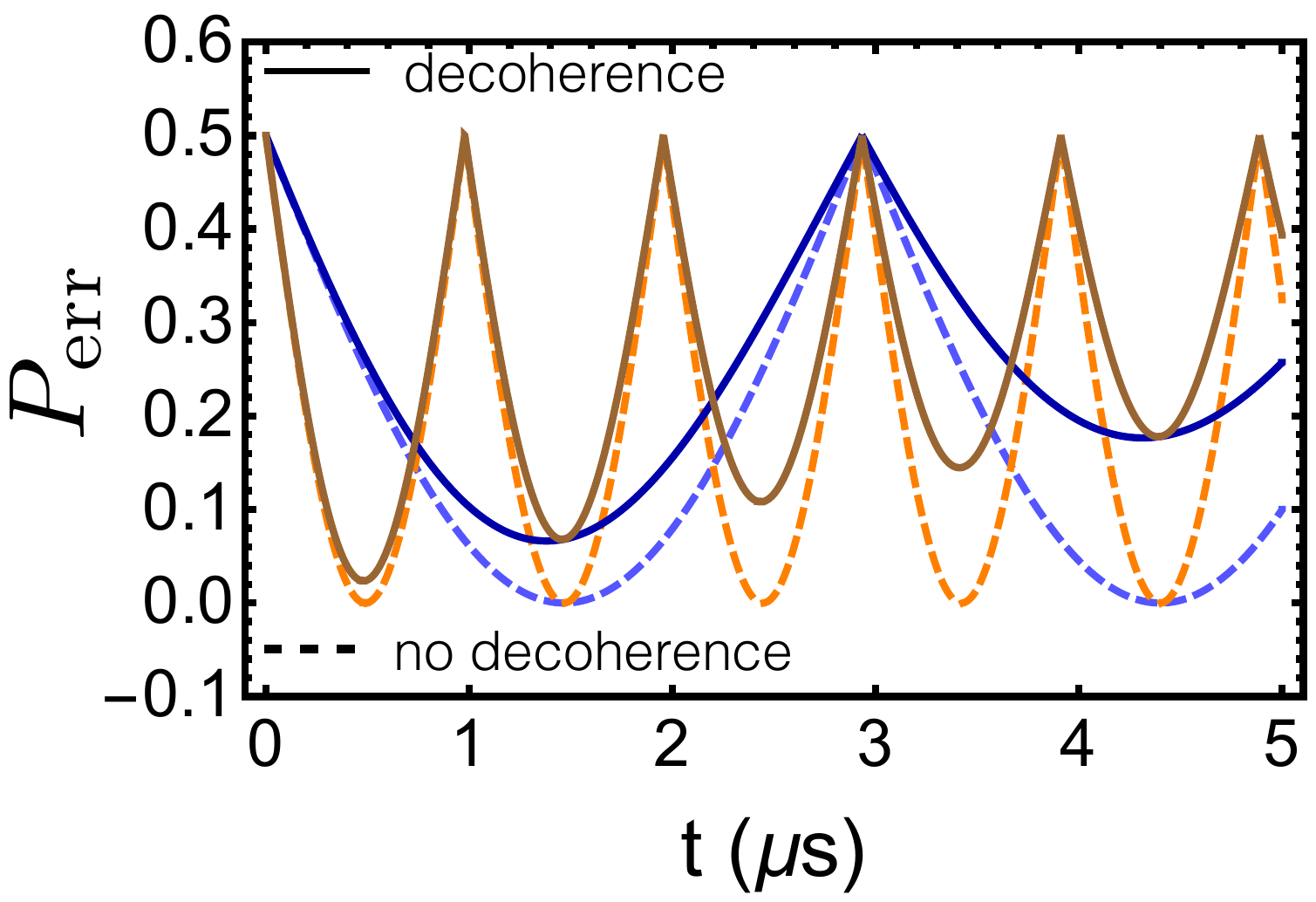}
        \caption[]
{Minimum error probability, $P_{\text{err}} $, with ($T_2 = 10$ $\mu$s) and without decoherence for two electric fields. Light/dark blue curves:  $\Delta E_x = 10^6$ V/m. Orange/brown curves:  $\Delta E_x = 3\times 10^6$ V/m. $E_{y,0} = 0 = \Delta E_y$, $B_z = 0$ for all curves.}\label{fig:tmin} 
        \end{centering}
\end{figure}
Calculation yields
\begin{equation}
\rho =\left(
\begin{array}{cc}
  \cos ^2\left({|\mathcal{E}| t}/{\hbar}\right) & ({i \hat{\mathcal{E}^*} /2 )\sin
   \left(2 {|\mathcal{E}| t}/{\hbar}\right)}{} \\
 ({-i\hat{\mathcal{E}}/{2
  }) \sin \left(2 {|\mathcal{E}| t}/{\hbar}\right)} &  \sin ^2\left({|\mathcal{E}| t}/{\hbar}\right). \\
\end{array}
\right)
\end{equation}
where $\hat{\mathcal{E}} = \mathcal{E}/|\mathcal{E}|$.
If $\mathcal{E}_0 || \mathcal{E}_1$ the eigenvalues of $\hat{\bm{\Lambda}}$ reduce to a particularly simple form:
\begin{equation}
\lambda_{\pm} = \frac{1}{2}(P_1-P_0) \pm \frac{1}{2}\sqrt{P_0^2 + P_1^2 - 2 P_0 P_1 \cos\left(  \frac{2 |\Delta \mathcal{E}|}{\hbar} t \right)  },
\end{equation}
where we find that the eigenvalues depend only on the perturbing electric field.
This fact is not true in general as seen in succeeding examples and discussed in Appendices \ref{appendixC} and \ref{appendixD}.
If $P_0 = P_1 = 1/2$ and $\mathcal{E}_0 || \mathcal{E}_1$,
\begin{equation}
\lambda_{\pm} = \pm \frac{1}{2}    |   \sin\left(  \frac{ |\Delta \mathcal{E}| }{\hbar} t \right)|.
\end{equation}
The 
eigenvector of the negative (positive) eigenvalue feeds in to $\hat{\bm{\Pi}}_0$ ($\hat{\bm{\Pi}}_1$) according to Eq.~(\ref{eq:POVM}).
Two conclusions are reached: (1) $P_{\text{err}}$ may vanish (at some times called $t_{min}$) only if the non-axial perturbing electric field is parallel to the unperturbed field, and (2) under the same conditions, $P_{\text{err}}$ will depend only upon the perturbing field.

The dashed curves of Figure \ref{fig:tmin} show $P_{err}$ evolving in time when the perturbing field is parallel to the unperturbed field ($\hat{x}$). In both instances shown the error probability falls to zero at each minima.

\subsection{Calculation of Minimum Error Probability: No Decoherence}

When an axial magnetic field is switched on, the solution to Eq. \ref{eq:SLE} is
\begin{equation}\label{}
\rho = \left(
\begin{array}{cc}
\rho_{11} & \rho_{12}  \\
\rho_{21} &  \rho_{22}\\
\end{array}
\right)
\end{equation}
with matrix elements
\begin{equation}
\rho_{11} =  \frac{1}{\mathcal{B}_z^2+|\mathcal{E}|^2}  \Bigg[\mathcal{B}_z^2+ |\mathcal{E}|^2 \cos^2 \left(\frac{\sqrt{\mathcal{B}_z^2+|\mathcal{E}|^2} t}{\hbar}\right) \Bigg],
\end{equation}
\begin{equation}
\rho_{22} =  1 - \rho_{11},
\end{equation}
and
\begin{widetext}
\begin{equation}
\rho_{21} = \frac{1}{2} \frac{\mathcal{E}}{\mathcal{B}_z^2+|\mathcal{E}|^2} \Bigg[  \mathcal{B}_z - \mathcal{B}_z \cos
   \left(2\frac{\sqrt{\mathcal{B}_z^2+|\mathcal{E}|^2} t}{\hbar}\right)  - \frac{i}{2} \sqrt{\mathcal{B}_z^2+|\mathcal{E}|^2} \sin
   \left(2\frac{\sqrt{\mathcal{B}_z^2+|\mathcal{E}|^2} t}{\hbar}\right)\Bigg], \qquad \rho_{12} = \rho_{21}^*,
\end{equation}
\end{widetext}
for arbitrary $\mathcal{E}$.
Incorporating the axial magnetic field reduces the symmetry and thwarts a simple reduction of the eigenvalues to functions of $\Delta \mathcal{E}$ which was possible in the previous case.

\begin{figure}[ptbh]
 \begin{centering}
        \includegraphics[scale = 0.26,trim = 15 120 0 50, angle = -0,clip]{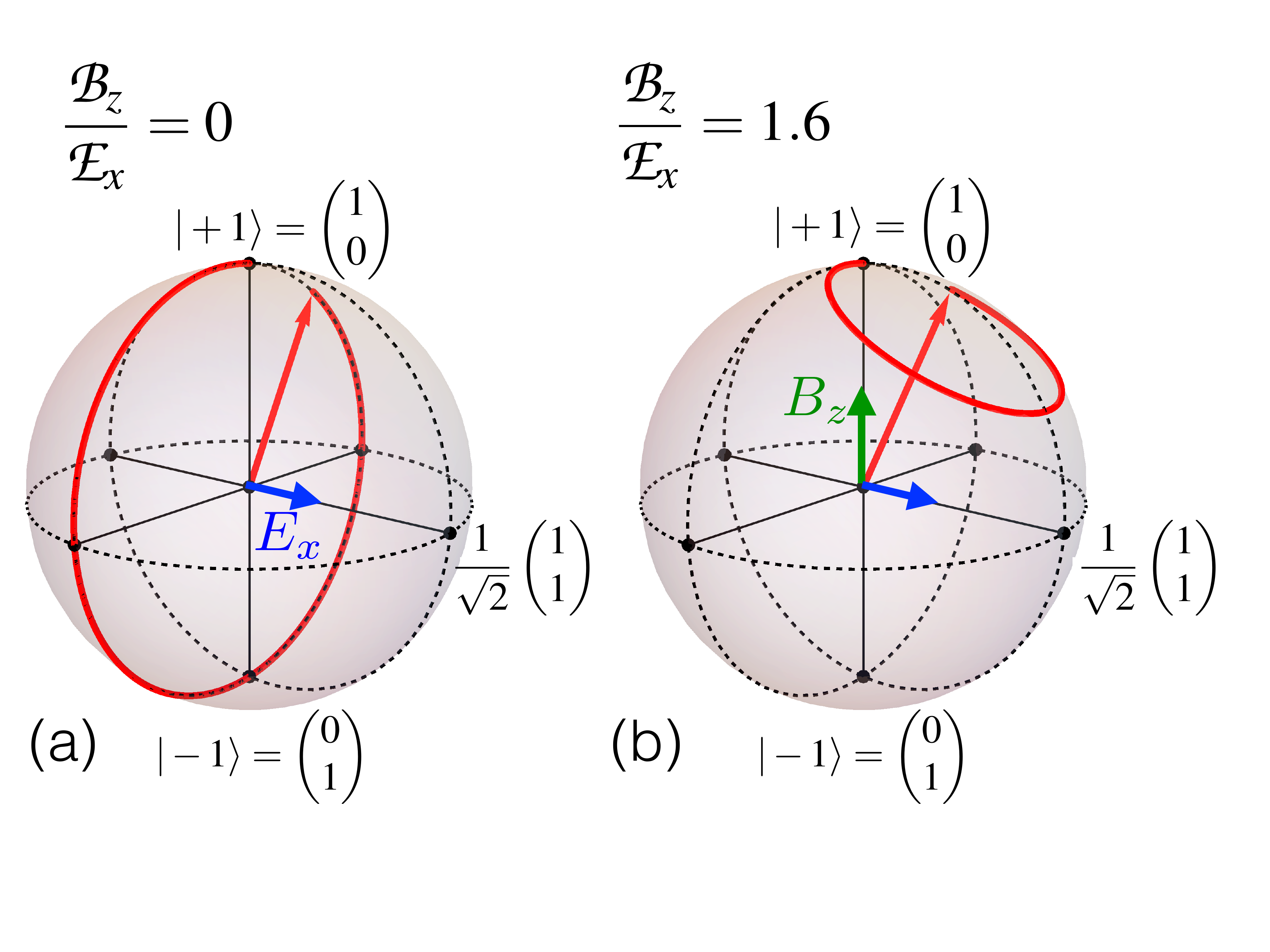}
        \caption[]
{Representation of the spin (red arrow) trajectory (red trace) on a Bloch sphere when an electric field is applied perpendicular to the axial direction with no magnetic field (a) and with an axial magnetic field (b). The spin starts in the $| +1 \rangle $ state. Decoherence is not considered in this figure.
}\label{fig:blochSpheres} 
        \end{centering}
\end{figure}

The role of the axial magnetic field can be understood by considering the Bloch sphere of Figure \ref{fig:blochSpheres}.
For the initial spin state chosen ($| +1 \rangle$) a magnetic field applied axially is detrimental to precise detection.
Precise discrimination of $\rho_0$ and $\rho_1$ requires  those states to be orthogonal which in our case would be $|+1\rangle$ and $|-1 \rangle$. 
Contrasting Figure \ref{fig:blochSpheres}(a ,b) uncovers the influence of the axial magnetic field; an axial magnetic field limits excursions into $|-1\rangle$ and therefore increases the error probability.
In Appendix \ref{appendixB} we examine other options for $\rho(0)$ for which an axial magnetic field is beneficial.

\subsection{Calculation of Minimum Error Probability: Decoherence \& $\mathcal{B}_z = 0$}\label{section:case3}

Decoherence may have two contributions: axial magnetic field noise and electric field noise. Which of these is dominant depends on the relative magnitudes of each field \cite{Dolde2011}.
If $\mathcal{B}_z \ll   |\mathcal{E}|$, there is no first order energy shift due to the magnetic field.  The spin state oscillates between the $|+1\rangle$ and $|-1\rangle$ states (Fig.~\ref{fig:blochSpheres}(a)) and  electric field noise parallel to the direction of the static electric field will be be the primary source of decoherence. The characteristics of the electric noise will depend on the electric source --- we model it phenomenologically via the decoherence rate $\kappa= T_2^{-1}$. $T_2$ is of the order of microseconds \cite{Balasubramanian2009}. Opposite reasoning applies when $\mathcal{B}_z \gg |\mathcal{E}|$; here decoherence is supplied by surface magnetic impurities with spins that fluctuate along the NV axis \cite{Rosskopf2014, Ajisaka2016}.
Either fluctuation is modeled as white noise (Markovian such that fluctuating field correlation time $\tau_c \rightarrow 0$) and treated with the Lindblad formalism \cite{Szankowski2013}.
Spin flip processes are inefficient and therefore ignored; the lack of these processes preserves the NV spin in the $ \{|+1\rangle, ~|-1\rangle\}$ basis such that the $|0 \rangle$ state can be safely avoided.

For an arbitrary transverse static electric field the Lindblad decoherence operator is
\begin{eqnarray}
 \hat{L}_d = 
\left(\frac{\kappa}{2}\right)^{1/2}
\left( {\begin{array}{cc}
   0 & \hat{\mathcal{E}^*}  \\
   \hat{\mathcal{E}}  & 0 \\
  \end{array} } \right),\nonumber
\end{eqnarray}
which describes decoherence for electric fluctuations along $\mathcal{E}$ (axial electric noise does not contribute to decoherence).
The Lindblad operator reduces to  $\hat{L}_d = \sqrt{\kappa/2}~ \sigma_x$ if $E_y$ and $\hat{L}_d = \sqrt{\kappa/2} ~\sigma_y$ if $E_x  = 0$.

Eq. \ref{eq:SLE} is solved to be
\begin{widetext}
\begin{equation}\label{}
\rho= \frac{1}{2}  \left(
\begin{array}{cc}
1+e^{-\kappa t}  \cos \left(2{|\mathcal{E}| t}/{\hbar}\right) & i \hat{\mathcal{E}^*} e^{-\kappa t} \sin
   \left(2{|\mathcal{E}| t}/{\hbar}\right) \\
-i \hat{\mathcal{E}} e^{-\kappa t} \sin
   \left(2{|\mathcal{E}| t}/{\hbar}\right) &  1 - e^{-\kappa t}  \cos \left(2{|\mathcal{E}| t}/{\hbar}\right) \\
\end{array}
\right)
\end{equation}
\end{widetext}
with $\mathcal{B}_z = 0$.
Since $B_z$ is negligible the two pertinent density matrices are 
\begin{widetext}
\begin{equation}\label{eq:rho0}
\rho_0 = \frac{1}{2}  \left(
\begin{array}{cc}
1+e^{-\kappa t}  \cos \left(2{|\mathcal{E}_0| t}/{\hbar}\right) & i \hat{\mathcal{E}_0^*}e^{-\kappa t} \sin
   \left(2{|\mathcal{E}_0| t}/{\hbar}\right) \\
-i \hat{\mathcal{E}_0}e^{-\kappa t} \sin
   \left(2{|\mathcal{E}_0| t}/{\hbar}\right) &  1 - e^{-\kappa t}  \cos \left(2{|\mathcal{E}_0| t}/{\hbar}\right) \\
\end{array}
\right),
\end{equation}
\begin{equation}\label{eq:rho1}
\rho_1 = \frac{1}{2}  \left(
\begin{array}{cc}
1+e^{-\kappa t}  \cos \left(2{|\mathcal{E}_0+ \Delta \mathcal{E}| t}/{\hbar}\right) & i \frac{(\mathcal{E}_0^*+ \Delta \mathcal{E}^*)}{|\mathcal{E}_0 + \Delta \mathcal{E}|} e^{-\kappa t} \sin
   \left(2{|\mathcal{E}_0 + \Delta \mathcal{E}| t}/{\hbar}\right) \\
-i \frac{(\mathcal{E}_0 + \Delta \mathcal{E})}{|\mathcal{E}_0 + \Delta \mathcal{E}|} e^{-\kappa t} \sin
   \left(2{|\mathcal{E}_0 + \Delta \mathcal{E}| t}/{\hbar}\right) &  1 - e^{-\kappa t}  \cos \left(2{|\mathcal{E}_0+ \Delta \mathcal{E}| t}/{\hbar}\right) \\
\end{array}
\right)
\end{equation}
\end{widetext}
Analysis of the error probability follows that of Section \ref{sec:case1}: (1) $P_{\text{err}}$ minima are smallest only if the non-axial perturbing electric field is parallel to the unperturbed field ($P_{\text{err}}$  will not vanish at any time now due to decoherence) (2) under the same conditions, $P_{\text{err}}$ will depend only upon the perturbing field.

\subsection{Static $E$-field measurement}\label{section:staticE}

The theory of quantum state discrimination dictates that the preferred measurement basis is given by the kets $|\phi_+ \rangle$ and $|\phi_- \rangle$ in the span of $\{| +1 \rangle, |-1  \rangle \}$ as defined earlier.
However measurements take place in the basis $\{|0\rangle, |\pm 1 \rangle \}$ through spin-dependent fluorescence;
to be concrete, photon emission is more likely when the electron is found in $|0\rangle$ compared to the other two states.

By unitarily manipulating the optimal NV state through a series of optical pulses, measurement in the standard basis provides electric field detection. The steps taken should be: (1) at the desired time, $t_{min}$ (to be discussed in next section), apply a specified optical sequence which rotates $|\phi_+\rangle$ to $|+1\rangle$ (2) now apply a separate pulse to rotate the state from $|+1 \rangle$ to $|0\rangle$.
The zero-field splitting, as shown in Figure \ref{fig:NVsensor}, allows this last pulse to completely convert $+1$ to $0$ since different resonance conditions exist for $+1$ and $-1$ with 0.
(3) laser light at the appropriate wavelength  is used to excite the NV state after which fluorescence  (occurring $< 10$ ns later \cite{Collins1983, Hanzawa1997}) is taken to be a measurement of the electric field while absence of fluorescence is measurement of no electric field. The conclusions of the measurement process are summarized in Table II.

\begin{table}[h]
\centering
\begin{tabular}{l l l}
\hline
Measurement & Quantum state & Postdiction\\
\hline
\hline
$\Pi_0$ `click' &  state is $\rho_0$  & E absent \\
\hline
$\Pi_1$ `click' &  state is $\rho_1$   & E present \\
\hline
photon emission &  state is $|0\rangle$   & E present \\
\hline\hline
\end{tabular}
\caption{\label{quanttab} Postdictions based on measurement outcomes.}
\end{table}

\section{Detection Optimization}\label{section:optimization}

For the initial density matrix of Eq.~(\ref{eq:init}),  an axial magnetic field always increases $P_{err}$.
Our first proposed strategy to optimize the detection is to use axial magnetic fields as small as possible  (and no transverse magnetic fields).
By neglecting the axial magnetic field we may use the results of Section \ref{section:case3}.

A second optimization strategy is to  bdevise a detector and sensor architecture where the perturbing electric field lies in the plane of the initial electric field and axial direction. To be concrete we set $E_{y,0} = 0$ so $\mathcal{E}_0 = \mathcal{E}_{x,0}$. Therefore we desire $\bm{E}_1 = E_{x,0}\hat{x} + \Delta E_x \hat{x} + \Delta E_z \hat{z}$ (we allow for a axial electric field since the spin is insensitive to it).
Equations (\ref{eq:rho0}) and (\ref{eq:rho1}) give us
\begin{widetext}
\begin{equation}\label{eq:rho0s}
\rho_0 = \frac{1}{2}  \left(
\begin{array}{cc}
1+e^{-\kappa t}  \cos \left(2{|\mathcal{E}_{x,0}| t}/{\hbar}\right) & i e^{-\kappa t} \sin
   \left(2 {|\mathcal{E}_{x,0}| t}/{\hbar}\right) \\
-i e^{-\kappa t} \sin
   \left(2 {|\mathcal{E}_{x,0}| t}/{\hbar}\right) &  1 - e^{-\kappa t}  \cos \left(2{|\mathcal{E}_{x,0}| t}/{\hbar}\right) \\
\end{array}
\right),
\end{equation}
\begin{equation}\label{eq:rho1s}
\rho_1 = \frac{1}{2}  \left(
\begin{array}{cc}
1+e^{-\kappa t}  \cos \left(2{|\mathcal{E}_{x,0} + \Delta \mathcal{E}_x| t}/{\hbar}\right) & i e^{-\kappa t} \sin
   \left(2 {|\mathcal{E}_{x,0} + \Delta \mathcal{E}_x| t}/{\hbar}\right) \\
-i e^{-\kappa t} \sin
   \left(2 {|\mathcal{E}_{x,0} + \Delta \mathcal{E}_x| t}/{\hbar}\right) &  1 - e^{-\kappa t}  \cos \left(2{|\mathcal{E}_{x,0} + \Delta \mathcal{E}_x| t}/{\hbar}\right) \\
\end{array}
\right).
\end{equation}
\end{widetext}
The eigenvalues of $\bm{\Lambda}$ are
\begin{equation}
\lambda_{\pm} = \frac{1}{2}(P_1-P_0) \pm  \frac{e^{-\kappa t}}{2}\sqrt{P_0^2 + P_1^2 - 2 P_0 P_1 \cos\left(  \frac{2 \Delta\mathcal{E}_x}{\hbar} t \right)  },
\end{equation}
or if $P_0 = P_1 = 1/2$:
\begin{equation}
\lambda_{\pm} = \pm  \frac{e^{-\kappa t}}{2} |   \sin\left(  \frac{\Delta\mathcal{E}_x }{\hbar} t \right)|
\end{equation}
such that
\begin{equation}\label{eq:POVMPE}
P_{\text{err}}  = \frac{1}{2} (1 - e^{-\kappa t} |\sin (\frac{\Delta\mathcal{E}_x }{\hbar}t) | ) 
\end{equation}

\begin{figure}[ptbh]
 \begin{centering}
        \includegraphics[width=\columnwidth]{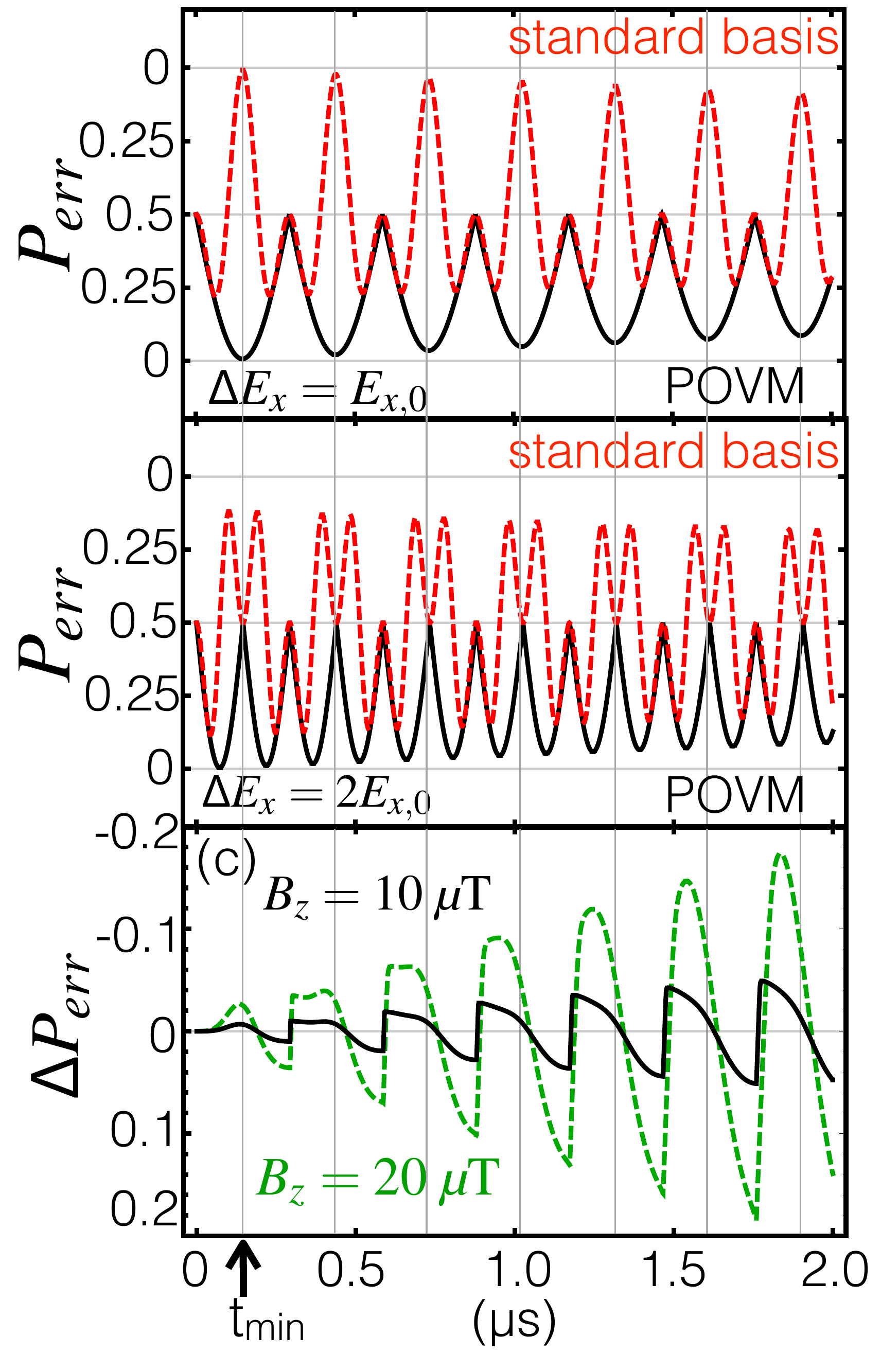}
        \caption[]
{(a, b) Error probability calculated using POVMs (black solid line using Eq. \ref{eq:POVMPE}) and the standard measurement basis (blue dashed line using Eq. \ref{eq:standardPE}) for electric fields $E_{x,0} = 10^7$ V/m and $E_{x,1}  = E_{x,0}  + \Delta E_{x}$. 
(a) $\Delta E_x$ is odd multiple of $E_{x,0}$: for certain times, the minimum $P_{err}$ in the standard basis is equivalent to the minimum $P_{err}$ in the POVM calculation.
 Vertical gray lines denote local minima in the error probability.
(b) $\Delta E_x$ is even multiple of $E_{x,0}$: the minimum $P_{err}$ in the standard basis is significantly worse than the minimum $P_{err}$ in the POVM calculation.
(c) Difference in $P_{err}$ when an axial magnetic field is applied: $\Delta P_e = P_{err}(B_z ) - P_{err}( 0)$ using the POVM calculation and the same electric fields as in (a).
The arrow on the bottom left of (c) points to $t_{min}$ corresponding to (a) --- the time at which the minimum error probability exists so the optimal time for a measurement to take place.
}\label{fig:ErrorProb} 
        \end{centering}
\end{figure}

If this is the case, the condition for minimal $P_{\text{err}} $ is accurately given by
\begin{equation}\label{eq:tmin}
t_{min} = n \frac{{\pi} \hbar}{2|\Delta \mathcal{E}_x|}, \qquad n = 1, 2, 3, ...
\end{equation}
when $\kappa \ll \Delta E_x/\hbar$.
The effects of sizable magnetic fields cannot be reduced to such a simple form and, as already mentioned, are to be avoided as they have a deleterious effect on $P_{\text{err}} $ for the initial spin chosen thus far.
A small axial magnetic field does not drastically alter the condition for minima as Figure \ref{fig:ErrorProb}(c) demonstrates (cf. black solid line for $B_z = 10 ~ \mu$T and green dotted line for $B_z = 20 ~ \mu$T).
One should understand the minima as times when a spin evolving in $\bm{E}_0$ is in a state most nearly orthogonal to the state of a spin evolving in $\bm{E}_1$. If there were no decoherence the states would be exactly orthogonal and  perfect discrimination would be possible at $t_{min}$.
Optimal detection is accomplished by taking measurements at $t_{min}$.

Since the state of the NV spin is only measurable in the standard basis, a sequence of unitary pulses must be applied to the spin in order to transform to the standard basis as discussed in Section \ref{section:staticE}. One may wonder how $P_{\text{err}} $ is affected by a simpler measurement where a $\pi/2$ pulse takes  $| -1 \rangle \rightarrow |0 \rangle$. The photo-luminescence of $|0 \rangle$ would signal an electric field.
$P_{\text{err}} $ in this reduced scheme (before the $\pi/2$ pulse) is 
\begin{equation}
P_{\text{err}}  = \frac{1}{2}\text{Tr}\left[ \left(
\begin{array}{cc}
1  & 0 \\
0&  0  \\
\end{array}
\right)\rho_1\right] + \frac{1}{2} \text{Tr}\left[ \left(
\begin{array}{cc}
0  & 0 \\
0&  1    \\
\end{array}
\right)\rho_0 \right]
\end{equation}
which simplifies to
\begin{equation}\label{eq:standardPE}
P_{\text{err}}  = \frac{1}{2} + \frac{1}{4} e^{-\kappa t} \left[\cos\left(\frac{2 (\mathcal{E}_{x,0} + \Delta \mathcal{E}_{x,0})}{\hbar}t\right) - \cos\left(\frac{2 \mathcal{E}_{x,0}}{\hbar}t\right) \right]
\end{equation}
where the density matrices are derived in Eqs.~(\ref{eq:rho0s}) and (\ref{eq:rho1s}).
Unlike with POVMs, the dependence of the error probability  on electric field cannot be written solely in terms of $\Delta E_x$.
From Eq.~(\ref{eq:standardPE}), it can be demonstrated (as in Figure \ref{fig:ErrorProb}(a)) that $P_{err}(t_{min})$ are equal between the two measurement bases as long as decoherence is not too strong \emph{and} $\Delta E_x = n E_{x,0}$ where $n$ is an odd positive integer.
Figure \ref{fig:ErrorProb} compares the two probabilities (black solid and blue solid lines).
The agreement between the two measurement choices is worst when $\Delta E_x = n E_{x,0}$ for $n$  an even positive integer as depicted in Figure \ref{fig:ErrorProb}(b), where the POVM $P_{err}(t_{min})$ is much less than the standard basis $P_{err}(t_{min})$.
 Therefore the standard basis is unlikely to be chosen as a measurement basis unless $E_{x,0} = 0$.

\begin{figure}[ptbh]
 \begin{centering}
        \includegraphics[width=\columnwidth]{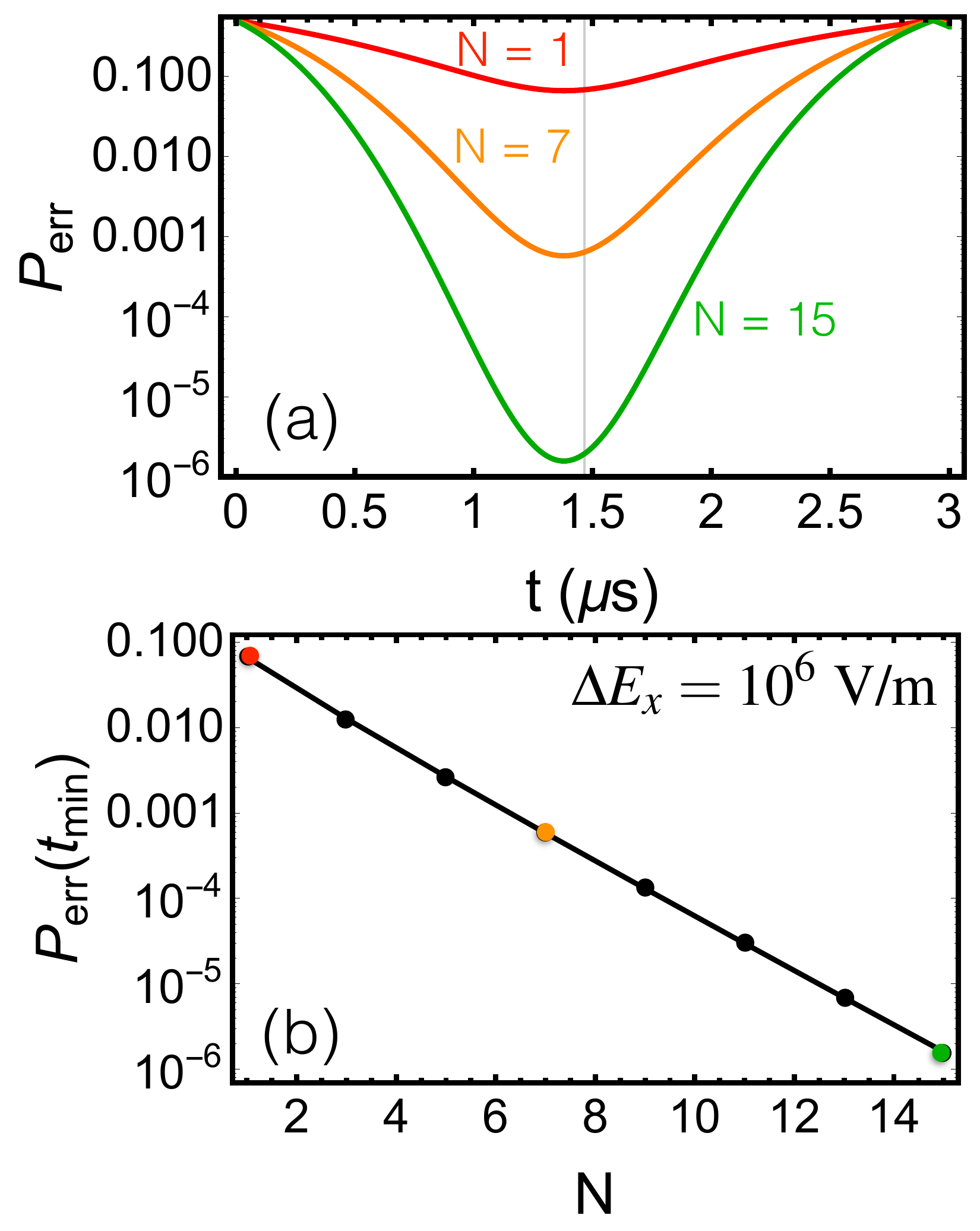}
        \caption[]
{
(a) Error probability, $P_{\text{err}} $, in time for $N = 1, 7, 15$ (red, orange, black) number of independent NV spin sensors. Vertical gray line denotes $t_{min}$ of Eq. (\ref{eq:tmin}); decoherence alters the actual time of $t_{min}$.
(b) Error probability, $P_{\text{err}} $, evaluated at $t_{min}$ for increasing number of independent NV spin sensors. Symbols are result of calculations; line is guide to eye. 
}\label{fig:NNVs} 
        \end{centering}
\end{figure}

By considering multiple independent defects in diamond, the error can be dramatically suppressed \cite{Happ2008,Higgins2011, Chaudhry2015, Wang2017}.
If greater than half of the spins suggest an electric field then an electric field is predicted with minimum error probability:
\begin{eqnarray}
P_{\text{err},N} &&= P_0 \sum_{n = 0}^{\lfloor N/2 \rfloor} {N \choose n} (1 - \text{Tr}(\rho_0 \Pi_1))^n  \text{Tr}(\rho_0 \Pi_1)^{N-n} \nonumber\\
{}&+& P_1 \sum_{n = 0}^{\lfloor N/2 \rfloor} {N \choose n} (1 - \text{Tr}(\rho_1 \Pi_0))^n  \text{Tr}(\rho_1 \Pi_0)^{N-n}
\end{eqnarray}
for odd $N$ while $P_{\text{err},N}  = P_{\text{err},N-1}$ for even N. 
Figure \ref{fig:NNVs} displays how $P_{\text{err}} $ changes as the number of independent sensors increases. 
$P_{\text{err}} $ falls off exponentially $P_{\text{err}}  \sim e^{-\alpha N}$ where in this particular example $\alpha \approx 0.75$.
Generally, the sensors are located at varying distances from the electric source and evolve under non-identical electric fields. 
The resulting probabilities $\text{Tr}(\rho_i \Pi_k)$ are different for each sensor and $P_{err,N}$ must be modified by replacing the binomial distributions with Poisson binomial distributions \cite{Wang1993}.
These more advanced considerations are beyond the scope of this paper.

\section{Detector Trade-Offs}\label{section:tradeoffs}

The minimum error probability is composed of two terms which are interpreted as either the dark count probability, $P_{dc}$ (the probability that the NV detects a photon when no photon was incident; also known as false positive probability) or the false negative probability, $P_{fn} $ (the probability that the NV does not detect a photon when one photon was incident).
\begin{equation}
P_{\text{err}} = P_0 P_{dc} + P_1 P_{fn} 
\end{equation}
with 
\begin{equation}
P_{dc} =  \text{Tr}(\rho_0 \hat{\Pi}_1) , \quad P_{fn} =   \text{Tr}(\rho_1 \hat{\Pi}_0).
\end{equation}
The dark count and false negative probabilities are equal if $P_0 = P_1$ (the POVMs depend on the \emph{a priori} probabilities in $\hat{\bm{\Lambda}}$).
In this case there is no trade-off between dark counts and false negatives.
 
\begin{figure}[ptbh]
 \begin{centering}
        \includegraphics[width=\columnwidth]{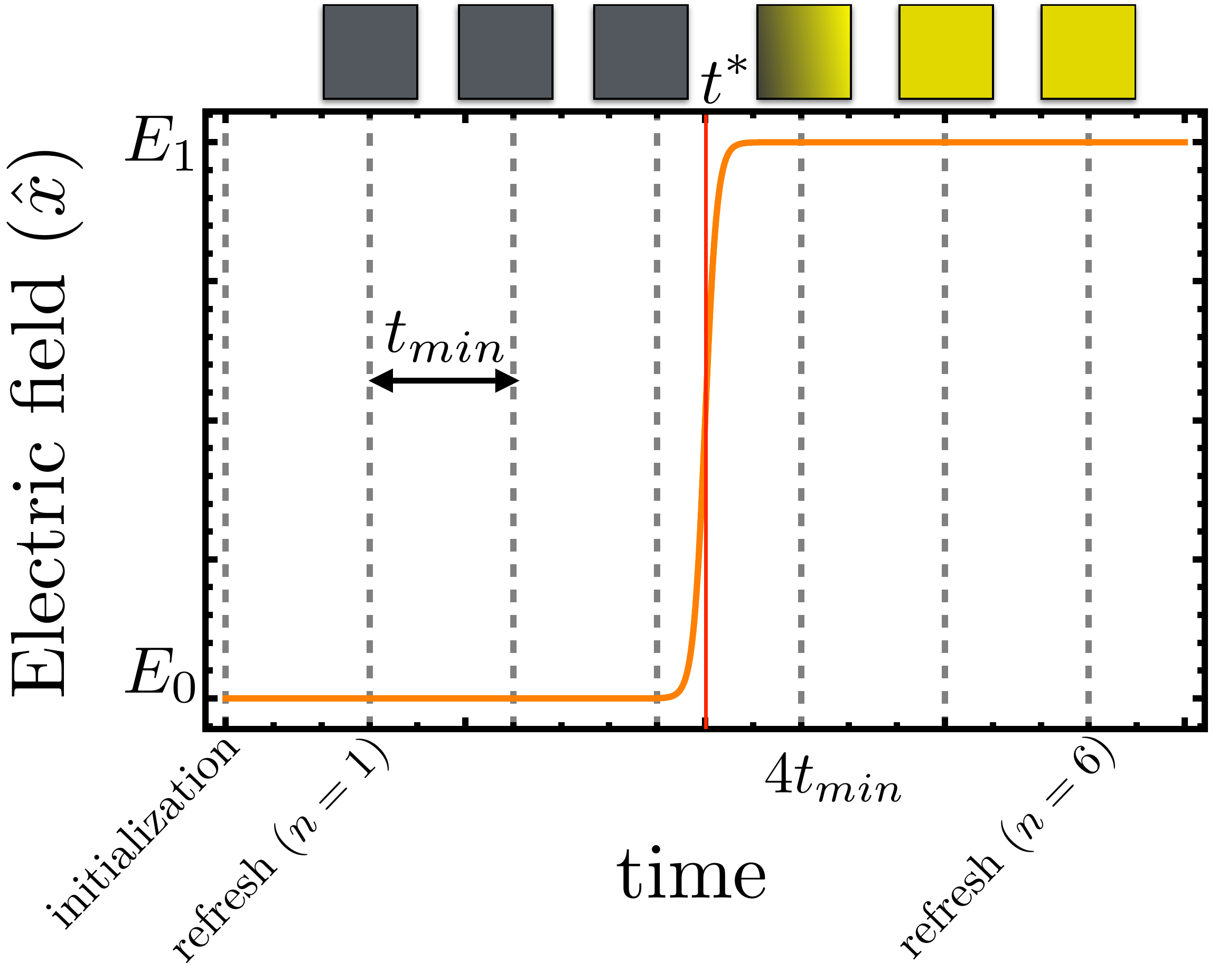}
        \caption[]
{
Measurement of $E$ turn-on time ($t^*$) protocol. After each measurement (at $n t_{min}$), the spin state is refreshed and reinitialized. Top boxes show outcome of measurements (dark or bright events). The measurement immediately after turn-on has larger error probability so is graded. Assuming small $P_{\text{err}} $ in all other measurements $t^*$ can be specified to within $2 t_{min}$. 
}\label{fig:turnon} 
        \end{centering}
\end{figure}

Up to now we have considered measurements of a static electric field. 
An operating detector should be able to distinguish the arrival time of the photon which is modeled as a sudden switch of an electric field from $E_{0}\hat{x}$ to $E_{1}\hat{x}$.
The goal is to be able to pin down a small time interval in which the electric field is turned on around time $t^*$. To accomplish this task, we propose that the method of static field detection, as introduced above, be applied in succession. If the process starts at $t = 0$, while $t < t^*$ and $P_{err} \ll 1$, measurements (at time $n t_{min}$ for $n$ iterations) will identify that there is no change in field due to an absorption process at the photoreceptive molecule. At times $t > t^* + t_{min}$ the measurement identifies that there is a change in the field, which is due to the absorption process of the photoreceptive molecule, and which Fig.~\ref{fig:turnon} demonstrates.
From the measurements we can deduce $t^*$ to within  $ 2 t_{min}$.

From Fig.~\ref{fig:turnon} we find that the uncertainty of the photon arrival time at the photoreceptive molecule (the  \emph{jitter})  is $2 t_{min}$ if $P_{\text{err}}$ is taken to be arbitrarily small (a limit realizable by adding and observing more NV sensors near the photoreceptive molecule). 
For larger $P_{\text{err}}$, jitter is $ > 2 t_{min}$ since multiple bright clicks would be needed to verify the photon arrival to a high degree of accuracy.
To decrease the jitter $t_{min}$ must be decreased which can only be accomplished by increasing the chromophore dipole field (see Eq.~(\ref{eq:tmin})) or changing to a spin center that is more sensitive to electric fields\cite{Tang2006}.
One could perform measurements at times $<t_{min}$ (thereby reducing jitter time) and these measurements may be accurate if a large number of sensors is used (see Figure \ref{fig:NNVs}). 

\section{Conclusion} 

We have described a general theoretical framework for the detection of single photons using a combination of a photoreceptive molecule and a quantum coherent spin center. This design allows the decoupling of the photon absorption event from the signal amplification process, by transduction of the photon arrival into a change in the local electric field environment and then detecting the effect of that change on the spin dynamics of the quantum coherent spin center. Use of multiple spin centers to independently detect the local electric field provides a dramatic reduction in the error rate of detection, as each spin evolves separately in the electric field of the activated photoreceptor. Although the framework for these measurements here has been assumed to be optical, and each spin would be independently measured, in principle this approach can be generalized to electrical measurements in which transport through the quantum spin provides information about the dynamics of the spin itself. We further note the potential of quantum spin centers whose coupling to electric fields is stronger than that of the NV center, even though the decoherence rates of such spins are likely to be much faster than those of the NV center.

\section{Acknowledgements} We acknowledge support of this work by DARPA/DETECT.

\section{Appendix} 

\appendix

\section{Positive Operator Valued Measurements (POVMs)} 

A set of Positive Operator Valued Measurements or POVMs possess the following three properties:
\begin{itemize}
    \item Hermitian: $\hat{\Pi}_n^{\dagger} = \hat{\Pi}_n$
    \item Positive: $\hat{\Pi}_n \ge 0$
    \item Complete: $\sum_n \hat{\Pi}_n =1$.
\end{itemize}
POVMs differ from projection operators by relaxing a fourth condition that the measurement operators be orthonormal ($\hat{\Pi}_m \hat{\Pi}_n = \hat{\Pi}_n \delta_{mn}$).

Any measurement can be described by a POVM and the relaxation of the orthonormality conditions allows an optimal set of POVMs to be determined that will minimize the error in the measurement. 
Assume that $N$ quantum states, specified by the set of $\rho_n$ density matrices are to be discriminated; that is, given a single measurement of the quantum system being $n$, the state of the system $\rho_n$ is predicted with some level of certainty. The probability that the state is correctly discriminated is 
\begin{equation}
P_{corr}  = \sum_{n=1}^N P_n \text{Tr}(\hat{\Pi}_n \rho_n)
\end{equation}
where $P_n$ is the \emph{a priori} probability of the system being in the $n$ state initially.
The POVMs chosen should minimize the error probability $P_{err} = 1 - P_{corr}$.
The minimization is described in Refs. \onlinecite{Helstrom1976, Holevo1982}. 

Discrimination between two states is particularly simple and the procedure to determine the minimum-error POVM is described in the main text. For more than two states there is no general form of the POVM minimizing error \cite{BarnettBook}.

\section{Eigenvalues of $\Lambda$ (no decoherence)}\label{appendixC}

Assume the defect is oriented in such a way that the baseline dipole field,  $\mathcal{E}_x  = d_{gs, \perp} E_x$, from the photorecepting molecule is along $\hat{x}$ only.
If no photon is absorbed the Hamiltonian of the spin ground state is $\mathscr{H}_0$ where
\begin{equation}
\mathscr{H}_{0}  = \left(
\begin{array}{cc}
\mathcal{D}   & \mathcal{E}_x   \\
\mathcal{E}_x  & \mathcal{D} \\
\end{array}
\right).
\end{equation}
Upon absorption of a single photon, the molecule reorients which leads to an altered electric field $\Delta\mathcal{E}_{\perp} =  d_{gs, \perp} (\Delta E_x + i \Delta E_y)$; the spin Hamiltonian in this case is $\mathscr{H}_0+\mathscr{H}_1$ where
\begin{equation}
\mathscr{H}_{1}  = \left(
\begin{array}{cc}
0   & \Delta \mathcal{E}_{\perp}^*  \\
 \Delta \mathcal{E}_{\perp}  & 0 \\
\end{array}
\right).
\end{equation}
We seek eigenvalues of $\Lambda = P_1 \rho_1 - P_0 \rho_0$ in order to determine the minimum error probability $P_e$.
When there is no decoherence the solutions to the Liouville equation are
\[
\rho_0 \equiv  \rho_0(t) =  e^{-i t \mathscr{H}_0/\hbar}
\rho(0)  e^{i t \mathscr{H}_0/\hbar},
\]
and
\[
\rho_1 \equiv  \rho_1(t) = e^{-i t (\mathscr{H}_0+\mathscr{H}_1)/\hbar}
\rho(0)  e^{-i t (\mathscr{H}_0+\mathscr{H}_1)/\hbar},
\]
$\rho_0(0) =\rho_1(0) \equiv \rho(0) $.
So 
\begin{widetext}
\begin{equation}
\Lambda = P_1  e^{-i t (\mathscr{H}_0+\mathscr{H}_1)/\hbar}
\rho(0)  e^{-i t (\mathscr{H}_0+\mathscr{H}_1)/\hbar} - P_0  e^{-i t \mathscr{H}_0/\hbar}
\rho(0)  e^{i t \mathscr{H}_0/\hbar}
\end{equation}
If $[\mathscr{H}_0, \mathscr{H}_1 ]= 0$ we can write
\begin{equation}
\Lambda = P_1  e^{-i t \mathscr{H}_0/\hbar}e^{-i t \mathscr{H}_1/\hbar}
\rho(0)  e^{-i t \mathscr{H}_0/\hbar} e^{-i t\mathscr{H}_1/\hbar} - P_0  e^{-i t \mathscr{H}_0/\hbar}
\rho(0)  e^{i t \mathscr{H}_0/\hbar}
\end{equation}
\begin{equation}
\Lambda = e^{-i t \mathscr{H}_0/\hbar} \Big[ P_1  e^{-i t \mathscr{H}_1/\hbar}
\rho(0)   e^{-i t\mathscr{H}_1/\hbar} - P_0 
\rho(0)\Big]  e^{i t \mathscr{H}_0/\hbar}
\end{equation}
\end{widetext}
The exponential outside the square brackets represent a unitary transformation on the operator within the square brackets.
The eigenvalues are the same in either case; \emph{i.e.} the eigenvalues of 
\begin{equation}
\Big[ P_1 \frac{1}{2} e^{-i t \mathscr{H}_1/\hbar}
\rho(0)   e^{-i t\mathscr{H}_1/\hbar} - P_0  \frac{1}{2} 
\rho(0)\Big] 
\end{equation}
are equal to the eigenvalues of $\Lambda$ as written above.
The eigenvectors of the two operators are not the same and will depend on the baseline electric field.
This analysis has not included decoherence effects.

Hence we see that the eigenvalues do not depend on the baseline electric field $E_x$ but only on the difference in electric field $\Delta E_x$.
If there is some $\Delta E_y$ we cannot equate the eigenvalues since we cannot factor out $e^{\pm i t \mathscr{H}_0/\hbar}$ since $[\mathscr{H}_0, \mathscr{H}_1] \neq 0$.

\section{Eigenvalues of $\Lambda$ (decoherence)}\label{appendixD}

Assume the defect is oriented in such a way that the baseline dipole field,  $\mathcal{E}_x  = d_{gs, \perp} E_x$, from the photorecepting molecule is along $\hat{x}$ only.
If no photon is absorbed the Hamiltonian of the spin ground state is $\mathscr{H}_0$ where
\begin{equation}
\mathscr{H}_{0}  = \left(
\begin{array}{cc}
\mathcal{D}   & \mathcal{E}_x   \\
\mathcal{E}_x  & \mathcal{D} \\
\end{array}
\right).
\end{equation}
Upon absorption of a single photon, the molecule reorients which leads to an altered electric field $\Delta\mathcal{E}_{\perp} =  d_{gs, \perp} (\Delta E_x + i \Delta E_y)$; the spin Hamiltonian in this case is $\mathscr{H}_0+\mathscr{H}_1$ where
\begin{equation}
\mathscr{H}_{1}  = \left(
\begin{array}{cc}
0   & \Delta \mathcal{E}_{\perp}^*  \\
 \Delta \mathcal{E}_{\perp}  & 0 \\
\end{array}
\right).
\end{equation}
We seek eigenvalues of $\Lambda = P_1 \rho_1 - P_0 \rho_0$ in order to determine the minimum error probability $P_e$.
When there is no decoherence the solutions to the Liouville equation are
\[
\rho_0 \equiv  \rho_0(t) = e^{-i t \mathscr{H}_0/\hbar}
\rho(0)  e^{i t \mathscr{H}_0/\hbar},
\]
and
\[
\rho_1 \equiv  \rho_1(t) = e^{-i t (\mathscr{H}_0+\mathscr{H}_1)/\hbar}
\rho(0)  e^{-i t (\mathscr{H}_0+\mathscr{H}_1)/\hbar},
\]
$\rho_0(0) =\rho_1(0) \equiv \rho(0) $.

\begin{equation}
\frac{\partial \rho}{\partial t} = -\frac{i}{\hbar} [\mathscr{H}_{gs} , \rho] +    \hat{L}_{d} \rho  \hat{L}_{d} ^{T}  - \frac{1}{2} \{ \hat{L}_{d} ^{T}  \hat{L}_{d} , \rho \} ,
\end{equation}
\begin{eqnarray}
 \hat{L}_d = 
\sqrt{\frac{\kappa}{2}}\sigma_x,\nonumber
\end{eqnarray}
which reduces the Liouville equation to 
\begin{equation}
\frac{\partial \rho}{\partial t} = -\frac{i}{\hbar} [\mathscr{H}_{0} + \mathscr{H}_1, \rho] +   \frac{\kappa}{2} \sigma_x \rho  \sigma_x  - \frac{\kappa}{2} \rho.
\end{equation}

This equation is now written as a superoperator equation; the commutator superoperator is
\begin{equation}
\mathscr{L}_0 = \mathscr{H}_0  \otimes \mathbb{1} -   \mathbb{1}\otimes \mathscr{H}_0
\end{equation}
\begin{equation}
\mathscr{L}_1 = \mathscr{H}_1  \otimes \mathbb{1} -   \mathbb{1}\otimes \mathscr{H}_1.
\end{equation}
The superoperator for the dissipation can be deduced by inspection to be
\begin{equation}
\mathscr{L}_d =  \sigma_x  \otimes \sigma_x -   \mathbb{1} \otimes  \mathbb{1}
\end{equation}
which then yields the spin dynamical equation
\begin{equation}
\frac{\partial \rho}{\partial t} = \big( -\frac{i}{\hbar} \mathscr{L}_0  -\frac{i}{\hbar} \mathscr{L}_1 + \frac{\kappa}{2} \mathscr{L}_d \big) \rho
\end{equation}
with solution
\begin{equation}
\rho = e^{-\frac{i}{\hbar} \mathscr{L}_0 t-\frac{i}{\hbar} \mathscr{L}_1 t+ \frac{\kappa}{2} \mathscr{L}_d t}\rho(0).
\end{equation}

So 
\begin{widetext}
\begin{equation}
\Lambda = P_1 e^{-\frac{i}{\hbar} \mathscr{L}_0 t-\frac{i}{\hbar} \mathscr{L}_1 t+ \frac{\kappa}{2} \mathscr{L}_d t}\rho(0) - P_0  e^{-\frac{i}{\hbar} \mathscr{L}_0 t + \frac{\kappa}{2} \mathscr{L}_d t}\rho(0).
\end{equation}
If  the $\mathscr{L}$'s commute, then by the Baker-Campbell-Hausdorff formula 
\begin{equation}
\Lambda = P_1 e^{-\frac{i}{\hbar} \mathscr{L}_0 t} e^{-\frac{i}{\hbar} \mathscr{L}_1 t} e^{ \frac{\kappa}{2} \mathscr{L}_d t}\rho(0) - P_0  e^{-\frac{i}{\hbar} \mathscr{L}_0 t}  e^{ \frac{\kappa}{2} \mathscr{L}_d t} \rho(0),
\end{equation}
or
\begin{equation}
\Lambda = e^{-\frac{i}{\hbar} \mathscr{L}_0 t} e^{ \frac{\kappa}{2} \mathscr{L}_d t} \big[ P_1  e^{-\frac{i}{\hbar} \mathscr{L}_1 t} \rho(0) - P_0  \rho(0) \big].
\end{equation}
\end{widetext}
The first exponential outside the square brackets is a unitary transformation and so it does not alter the eigenvalues. The second exponential is dissipative and will affect eigenvalues.
Hence the eigenvalues of $\Lambda$ will only depend on $\Delta E_x$ when there is no $E_y$ field.
The eigenvectors, however, are affected by $E_x$.

Hence we see that the eigenvalues do not depend on the baseline electric field $E_x$ but only on the difference in electric field $\Delta E_x$.
Again it was necessary that  $[\mathscr{H}_0, \mathscr{H}_1] = 0$ and $[\mathscr{H}_0, L_d] = 0$.

\section{Alternate sensor preparation}\label{appendixB}

Properties of the sensor change when the sensing spin is prepared in some other initial state -- for instance $\frac{1}{\sqrt{2}}(| +1 \rangle + |-1 \rangle)$.
Introducing an axial magnetic field, $B_z$,  may now favorably affect the sensor when the electric field to be detected is parallel to $\hat{x}$.
Conditions with a magnetic field may be necessary for state manipulation so protecting the sensor in a magnetic field is important.
\begin{figure}[ptbh]
 \begin{centering}
        \includegraphics[width=\columnwidth]{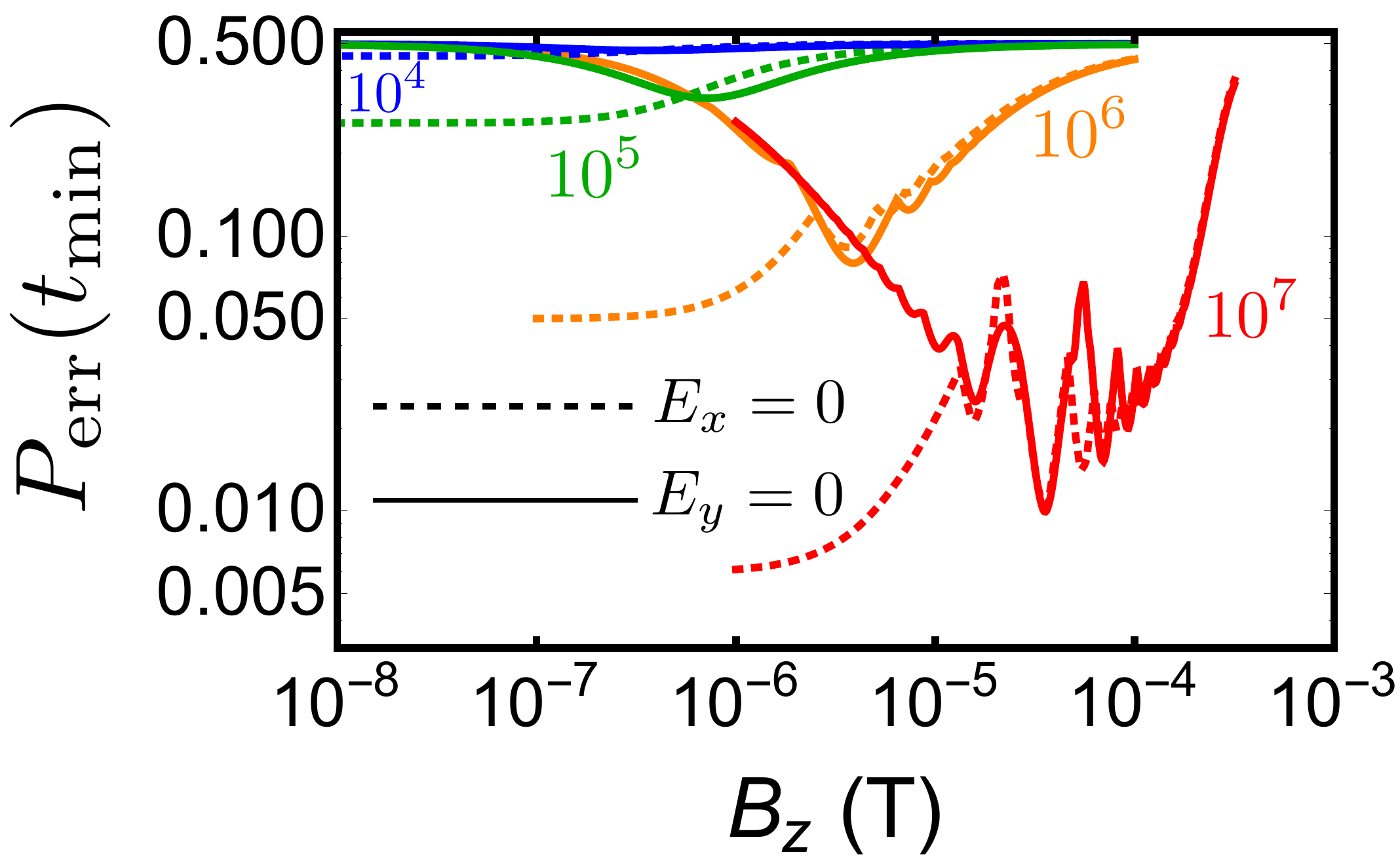}
        \caption[]
{Minimum error probability evaluated at the optimal time, $t_{min}$, as a function of axial magnetic field, $B_z$. Whether the electric field  (magnitude in V/m labels each colored curve) is oriented parallel (dotted, $E_x = 0$) or perpendicular (solid, $E_y = 0$) to the initial spin state dramatically affects the error probability.  Prior probabilities are $P_0 = P_1 = 1/2$. Decoherence is due to magnetic fluctuations along $\hat{z}$.
}\label{fig:PevB} 
        \end{centering}
\end{figure}
Figure~\ref{fig:PevB} displays $P_{\text{err}} $ versus the optimal axial magnetic field strength for $x-$ and $y-$directed electric fields ($E_0$ is taken to be zero).
It is apparent that for electric fields in $\hat{y}$ (dashed) the optimal magnetic field is a zero magnetic field, whereas a finite magnetic field distinguishes states better when the electric field is parallel to $\hat{x}$ (solid).
In either case, the crossover regime occurs at a $B_z$ roughly corresponding to $g\mu_B B_z \sim E_i d_{gs, \perp}$.

\begin{widetext}
\begin{figure*}[ptbh]
 \begin{centering}
        \includegraphics[scale = 0.35,trim = 0 0 0 0, angle = -0,clip]{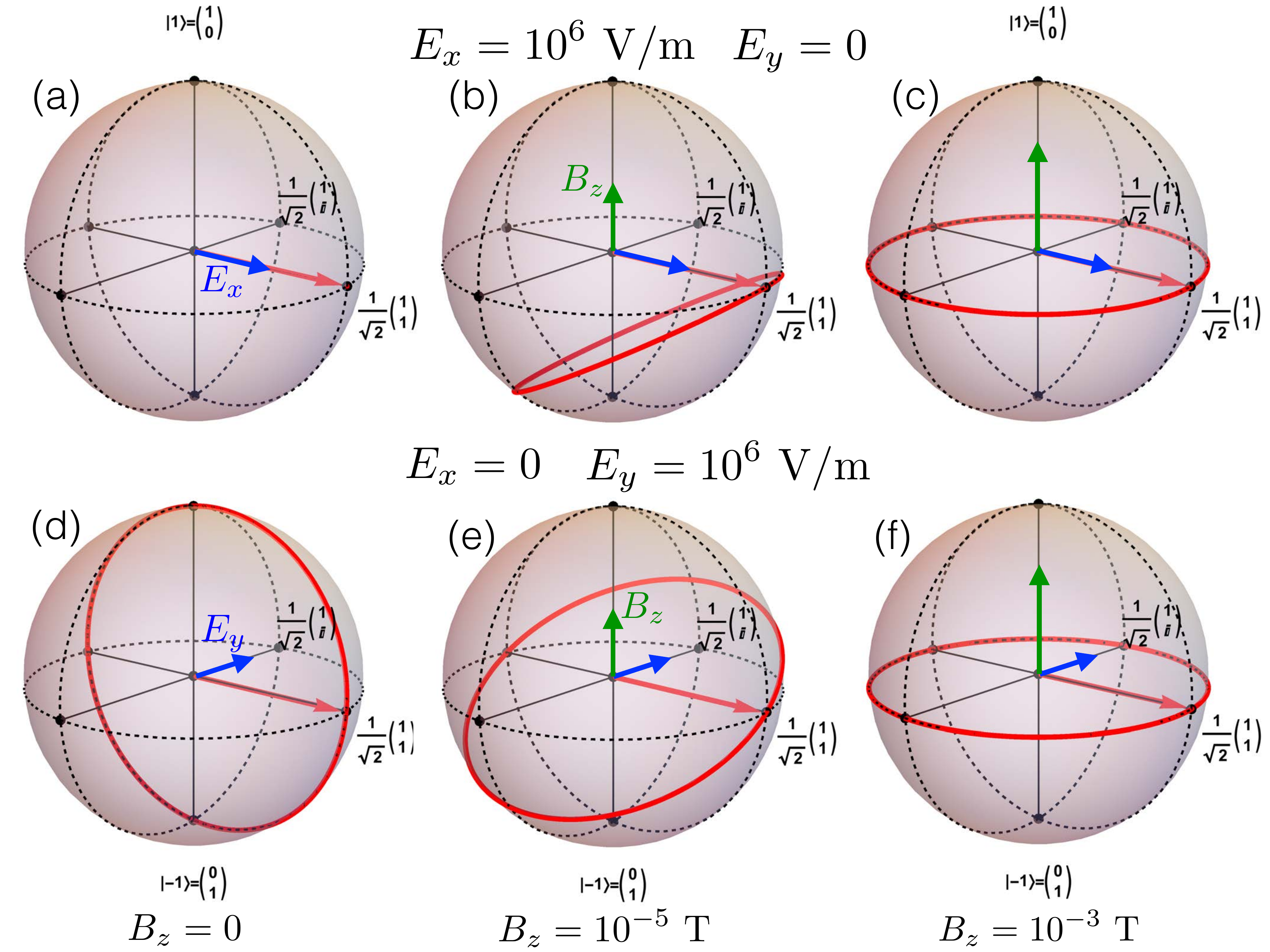}
        \caption[]
{Representation of the spin on a Bloch sphere for two different electric field configurations (by row) and three different magnetic fields (by column). Red arrow is initial spin state $2^{-1/2}(| +1 \rangle + |-1 \rangle)$. Red curve is trace of spin state in time. Decoherence is taken to be zero.
}\label{fig:blochSpheres2} 
        \end{centering}
\end{figure*}
\end{widetext}

Inspection of the ground state Hamiltonian clarifies why the electric fields contribute in this manner: the $i$-electric field, $E_i$, acts as a pseudomagnetic field in the $i$-direction.
If there is no $B_z$, then $E_x$ does nothing to the density matrix, as shown in Fig.~\ref{fig:blochSpheres2} (a) (solid curves in Figure \ref{fig:PevB}). In a small $B_z$, the spin is kicked away from $\hat{x}$, which allows $E_x$ to modify the density matrix (as a pseudomagnetic field in $\hat{x}$).
This will result in the spin evolving through states partially orthogonal to the initial state (Figure \ref{fig:blochSpheres2} (b)) which is  not possible without $E_x$. 
In this way the ability to distinguish the electric field is improved, so $P_{\text{err}} $ decreases. 
Now if $B_z$ gets much larger than the electric field precession, then the spin largely unaffected by the activated electric field  --- thus the error probability increases back to one half (Figure \ref{fig:blochSpheres2} (c)). 

Now consider the $y$-electric field, $E_y$, which acts like a magnetic field in the $y$-direction (dotted curves in Figure \ref{fig:PevB}).
Already at $B_z = 0$, this electric field distinguishes the electric field by rotating the spin into states orthogonal to the initial state as demonstrated in Figure \ref{fig:blochSpheres2} (d).
Similar to Figure \ref{fig:blochSpheres2} (c), if $B_z$ gets much larger than $E_y$, then the spin largely unaffected by the activated electric field and the error probability increases back to one half (Figure \ref{fig:blochSpheres2} (e,f)). 


\begin{thebibliography}{36}%
\makeatletter
\providecommand \@ifxundefined [1]{%
 \@ifx{#1\undefined}
}%
\providecommand \@ifnum [1]{%
 \ifnum #1\expandafter \@firstoftwo
 \else \expandafter \@secondoftwo
 \fi
}%
\providecommand \@ifx [1]{%
 \ifx #1\expandafter \@firstoftwo
 \else \expandafter \@secondoftwo
 \fi
}%
\providecommand \natexlab [1]{#1}%
\providecommand \enquote  [1]{``#1''}%
\providecommand \bibnamefont  [1]{#1}%
\providecommand \bibfnamefont [1]{#1}%
\providecommand \citenamefont [1]{#1}%
\providecommand \href@noop [0]{\@secondoftwo}%
\providecommand \href [0]{\begingroup \@sanitize@url \@href}%
\providecommand \@href[1]{\@@startlink{#1}\@@href}%
\providecommand \@@href[1]{\endgroup#1\@@endlink}%
\providecommand \@sanitize@url [0]{\catcode `\\12\catcode `\$12\catcode
  `\&12\catcode `\#12\catcode `\^12\catcode `\_12\catcode `\%12\relax}%
\providecommand \@@startlink[1]{}%
\providecommand \@@endlink[0]{}%
\providecommand \url  [0]{\begingroup\@sanitize@url \@url }%
\providecommand \@url [1]{\endgroup\@href {#1}{\urlprefix }}%
\providecommand \urlprefix  [0]{URL }%
\providecommand \Eprint [0]{\href }%
\providecommand \doibase [0]{http://dx.doi.org/}%
\providecommand \selectlanguage [0]{\@gobble}%
\providecommand \bibinfo  [0]{\@secondoftwo}%
\providecommand \bibfield  [0]{\@secondoftwo}%
\providecommand \translation [1]{[#1]}%
\providecommand \BibitemOpen [0]{}%
\providecommand \bibitemStop [0]{}%
\providecommand \bibitemNoStop [0]{.\EOS\space}%
\providecommand \EOS [0]{\spacefactor3000\relax}%
\providecommand \BibitemShut  [1]{\csname bibitem#1\endcsname}%
\let\auto@bib@innerbib\@empty
\bibitem [{\citenamefont {O'Connor}(1984)}]{OConnor1984}%
  \BibitemOpen
  \bibfield  {author} {\bibinfo {author} {\bibfnamefont {Desmond}\ \bibnamefont
  {O'Connor}},\ }\href@noop {} {\emph {\bibinfo {title} {Time-correlated
  single-photon counting}}}\ (\bibinfo  {publisher} {Elsevier},\ \bibinfo
  {address} {New York},\ \bibinfo {year} {1984})\BibitemShut {NoStop}%
\bibitem [{\citenamefont {Hadfield}(2009)}]{Hadfield2009}%
  \BibitemOpen
  \bibfield  {author} {\bibinfo {author} {\bibfnamefont {R.~H.}\ \bibnamefont
  {Hadfield}},\ }\bibfield  {title} {\enquote {\bibinfo {title} {Single-photon
  detectors for optical quantum information applications},}\ }\href@noop {}
  {\bibfield  {journal} {\bibinfo  {journal} {Nature Photonics}\ }\textbf
  {\bibinfo {volume} {3}},\ \bibinfo {pages} {696--705} (\bibinfo {year}
  {2009})}\BibitemShut {NoStop}%
\bibitem [{\citenamefont {McIntyre}(1966)}]{McIntyre1966}%
  \BibitemOpen
  \bibfield  {author} {\bibinfo {author} {\bibfnamefont {R.~J.}\ \bibnamefont
  {McIntyre}},\ }\bibfield  {title} {\enquote {\bibinfo {title} {Multiplication
  noise in uniform avalanche diodes},}\ }\href@noop {} {\bibfield  {journal}
  {\bibinfo  {journal} {IEEE Transactions on Electron Devices}\ }\textbf
  {\bibinfo {volume} {ED-13}},\ \bibinfo {pages} {164--168} (\bibinfo {year}
  {1966})}\BibitemShut {NoStop}%
\bibitem [{\citenamefont {Stillman}\ and\ \citenamefont
  {Wolfe}(1977)}]{Stillman1977}%
  \BibitemOpen
  \bibfield  {author} {\bibinfo {author} {\bibfnamefont {G.~E.}\ \bibnamefont
  {Stillman}}\ and\ \bibinfo {author} {\bibfnamefont {C.~M.}\ \bibnamefont
  {Wolfe}},\ }\enquote {\bibinfo {title} {Avalanche photodiodes},}\ in\ \href
  {\doibase 10.1016/S0080-8784(08)60150-7} {\emph {\bibinfo {booktitle}
  {Semiconductors and Semimetals}}},\ Vol.~\bibinfo {volume} {12}\ (\bibinfo
  {publisher} {Elsevier},\ \bibinfo {address} {New York},\ \bibinfo {year}
  {1977})\ Chap.\ \bibinfo {chapter} {chapter 5}, pp.\ \bibinfo {pages}
  {291--393}\BibitemShut {NoStop}%
\bibitem [{\citenamefont {Eisaman}\ \emph {et~al.}(2011)\citenamefont
  {Eisaman}, \citenamefont {Fan}, \citenamefont {Migdall},\ and\ \citenamefont
  {Polyakov}}]{Eisaman2011}%
  \BibitemOpen
  \bibfield  {author} {\bibinfo {author} {\bibfnamefont {M.~D.}\ \bibnamefont
  {Eisaman}}, \bibinfo {author} {\bibfnamefont {J.}~\bibnamefont {Fan}},
  \bibinfo {author} {\bibfnamefont {A.}~\bibnamefont {Migdall}}, \ and\
  \bibinfo {author} {\bibfnamefont {S.~V.}\ \bibnamefont {Polyakov}},\
  }\bibfield  {title} {\enquote {\bibinfo {title} {Invited review article:
  Single-photon sources and detectors},}\ }\href {\doibase 10.1063/1.3610677}
  {\bibfield  {journal} {\bibinfo  {journal} {Review of Scientific
  Instruments}\ }\textbf {\bibinfo {volume} {82}},\ \bibinfo {pages} {071101}
  (\bibinfo {year} {2011})},\ \Eprint
  {http://arxiv.org/abs/https://doi.org/10.1063/1.3610677}
  {https://doi.org/10.1063/1.3610677} \BibitemShut {NoStop}%
\bibitem [{\citenamefont {Natarajan}\ \emph {et~al.}(2012)\citenamefont
  {Natarajan}, \citenamefont {Tanner},\ and\ \citenamefont
  {Hadfield}}]{Natarajan2012}%
  \BibitemOpen
  \bibfield  {author} {\bibinfo {author} {\bibfnamefont {Chandra~M.}\
  \bibnamefont {Natarajan}}, \bibinfo {author} {\bibfnamefont {Michael~G.}\
  \bibnamefont {Tanner}}, \ and\ \bibinfo {author} {\bibfnamefont {Robert~H.}\
  \bibnamefont {Hadfield}},\ }\bibfield  {title} {\enquote {\bibinfo {title}
  {Superconducting nanowire single-photon detectors: physics and
  applications},}\ }\href@noop {} {\bibfield  {journal} {\bibinfo  {journal}
  {Superconductor Science and Technology}\ }\textbf {\bibinfo {volume} {25}},\
  \bibinfo {pages} {063001} (\bibinfo {year} {2012})}\BibitemShut {NoStop}%
\bibitem [{\citenamefont {Dolde}\ \emph {et~al.}(2011)\citenamefont {Dolde},
  \citenamefont {Fedder}, \citenamefont {Doherty}, \citenamefont
  {N{\"{o}}bauer}, \citenamefont {Rempp}, \citenamefont {Balasubramanian},
  \citenamefont {Wolf}, \citenamefont {Reinhard}, \citenamefont {Hollenberg},
  \citenamefont {Jelezko},\ and\ \citenamefont {Wrachtrup}}]{Dolde2011}%
  \BibitemOpen
  \bibfield  {author} {\bibinfo {author} {\bibfnamefont {F.}~\bibnamefont
  {Dolde}}, \bibinfo {author} {\bibfnamefont {H.}~\bibnamefont {Fedder}},
  \bibinfo {author} {\bibfnamefont {M.~W.}\ \bibnamefont {Doherty}}, \bibinfo
  {author} {\bibfnamefont {T.}~\bibnamefont {N{\"{o}}bauer}}, \bibinfo {author}
  {\bibfnamefont {F.}~\bibnamefont {Rempp}}, \bibinfo {author} {\bibfnamefont
  {G.}~\bibnamefont {Balasubramanian}}, \bibinfo {author} {\bibfnamefont
  {T.}~\bibnamefont {Wolf}}, \bibinfo {author} {\bibfnamefont {F.}~\bibnamefont
  {Reinhard}}, \bibinfo {author} {\bibfnamefont {L.~C.~L.}\ \bibnamefont
  {Hollenberg}}, \bibinfo {author} {\bibfnamefont {F.}~\bibnamefont {Jelezko}},
  \ and\ \bibinfo {author} {\bibfnamefont {J.}~\bibnamefont {Wrachtrup}},\
  }\bibfield  {title} {\enquote {\bibinfo {title} {{Electric-field sensing
  using single diamond spins}},}\ }\href {\doibase 10.1038/nphys1969}
  {\bibfield  {journal} {\bibinfo  {journal} {Nature Physics}\ }\textbf
  {\bibinfo {volume} {7}},\ \bibinfo {pages} {459--463} (\bibinfo {year}
  {2011})}\BibitemShut {NoStop}%
\bibitem [{\citenamefont {Tang}\ \emph {et~al.}(2006)\citenamefont {Tang},
  \citenamefont {Levy},\ and\ \citenamefont {Flatt\'e}}]{Tang2006}%
  \BibitemOpen
  \bibfield  {author} {\bibinfo {author} {\bibfnamefont {J.-M.}\ \bibnamefont
  {Tang}}, \bibinfo {author} {\bibfnamefont {J.}~\bibnamefont {Levy}}, \ and\
  \bibinfo {author} {\bibfnamefont {M.~E.}\ \bibnamefont {Flatt\'e}},\
  }\bibfield  {title} {\enquote {\bibinfo {title} {All-electrical control of
  single ion spins in a semiconductor},}\ }\href@noop {} {\bibfield  {journal}
  {\bibinfo  {journal} {\prl}\ }\textbf {\bibinfo {volume} {97}},\ \bibinfo
  {pages} {106803} (\bibinfo {year} {2006})}\BibitemShut {NoStop}%
\bibitem [{\citenamefont {Barnett}(2009)}]{BarnettBook}%
  \BibitemOpen
  \bibfield  {author} {\bibinfo {author} {\bibfnamefont {Stephen~M.}\
  \bibnamefont {Barnett}},\ }\href@noop {} {\emph {\bibinfo {title} {Quantum
  Information}}}\ (\bibinfo  {publisher} {Oxford University Press},\ \bibinfo
  {year} {2009})\BibitemShut {NoStop}%
\bibitem [{\citenamefont {Chaudhry}(2015)}]{Chaudhry2015}%
  \BibitemOpen
  \bibfield  {author} {\bibinfo {author} {\bibfnamefont {Adam~Zaman}\
  \bibnamefont {Chaudhry}},\ }\bibfield  {title} {\enquote {\bibinfo {title}
  {{Detecting the presence of weak magnetic fields using nitrogen-vacancy
  centers}},}\ }\href {\doibase 10.1103/PhysRevA.91.062111} {\bibfield
  {journal} {\bibinfo  {journal} {Physical Review A}\ }\textbf {\bibinfo
  {volume} {91}},\ \bibinfo {pages} {062111} (\bibinfo {year}
  {2015})}\BibitemShut {NoStop}%
\bibitem [{\citenamefont {Kastner}(1992)}]{Kastner1992}%
  \BibitemOpen
  \bibfield  {author} {\bibinfo {author} {\bibfnamefont {M.~A.}\ \bibnamefont
  {Kastner}},\ }\bibfield  {title} {\enquote {\bibinfo {title} {The
  single-electron transistor},}\ }\href {\doibase 10.1103/RevModPhys.64.849}
  {\bibfield  {journal} {\bibinfo  {journal} {Rev. Mod. Phys.}\ }\textbf
  {\bibinfo {volume} {64}},\ \bibinfo {pages} {849--858} (\bibinfo {year}
  {1992})}\BibitemShut {NoStop}%
\bibitem [{\citenamefont {Herzog}\ and\ \citenamefont
  {Bergou}(2004)}]{Herzog2004}%
  \BibitemOpen
  \bibfield  {author} {\bibinfo {author} {\bibfnamefont {Ulrike}\ \bibnamefont
  {Herzog}}\ and\ \bibinfo {author} {\bibfnamefont {J{\'{a}}nos~A.}\
  \bibnamefont {Bergou}},\ }\bibfield  {title} {\enquote {\bibinfo {title}
  {{Distinguishing mixed quantum states: Minimum-error discrimination versus
  optimum unambiguous discrimination}},}\ }\href {\doibase
  10.1103/PhysRevA.70.022302} {\bibfield  {journal} {\bibinfo  {journal}
  {Physical Review A}\ }\textbf {\bibinfo {volume} {70}},\ \bibinfo {pages}
  {1--6} (\bibinfo {year} {2004})}\BibitemShut {NoStop}%
\bibitem [{\citenamefont {Yu}\ and\ \citenamefont {Cardona}(2001)}]{Yu3ed}%
  \BibitemOpen
  \bibfield  {author} {\bibinfo {author} {\bibfnamefont {P.~Y.}\ \bibnamefont
  {Yu}}\ and\ \bibinfo {author} {\bibfnamefont {M.}~\bibnamefont {Cardona}},\
  }\href@noop {} {\emph {\bibinfo {title} {Fundamentals of semiconductors}}},\
  \bibinfo {edition} {3rd}\ ed.\ (\bibinfo  {publisher} {Springer-Verlag},\
  \bibinfo {address} {Berlin},\ \bibinfo {year} {2001})\BibitemShut {NoStop}%
\bibitem [{\citenamefont {Doherty}\ \emph {et~al.}(2013)\citenamefont
  {Doherty}, \citenamefont {Manson}, \citenamefont {Delaney}, \citenamefont
  {Jelezko}, \citenamefont {Wrachtrup},\ and\ \citenamefont
  {Hollenberg}}]{Doherty2013}%
  \BibitemOpen
  \bibfield  {author} {\bibinfo {author} {\bibfnamefont {Marcus~W.}\
  \bibnamefont {Doherty}}, \bibinfo {author} {\bibfnamefont {Neil~B.}\
  \bibnamefont {Manson}}, \bibinfo {author} {\bibfnamefont {Paul}\ \bibnamefont
  {Delaney}}, \bibinfo {author} {\bibfnamefont {Fedor}\ \bibnamefont
  {Jelezko}}, \bibinfo {author} {\bibfnamefont {J{\"{o}}rg}\ \bibnamefont
  {Wrachtrup}}, \ and\ \bibinfo {author} {\bibfnamefont {Lloyd~C.L.}\
  \bibnamefont {Hollenberg}},\ }\bibfield  {title} {\enquote {\bibinfo {title}
  {{The nitrogen-vacancy colour centre in diamond}},}\ }\href {\doibase
  10.1016/j.physrep.2013.02.001} {\bibfield  {journal} {\bibinfo  {journal}
  {Physics Reports}\ }\textbf {\bibinfo {volume} {528}},\ \bibinfo {pages}
  {1--45} (\bibinfo {year} {2013})}\BibitemShut {NoStop}%
\bibitem [{\citenamefont {Ajisaka}\ and\ \citenamefont
  {Band}(2016)}]{Ajisaka2016}%
  \BibitemOpen
  \bibfield  {author} {\bibinfo {author} {\bibfnamefont {Shigeru}\ \bibnamefont
  {Ajisaka}}\ and\ \bibinfo {author} {\bibfnamefont {Y.~B.}\ \bibnamefont
  {Band}},\ }\bibfield  {title} {\enquote {\bibinfo {title} {{Decoherence of
  three-level systems: Application to nitrogen-vacancy centers in diamond near
  a surface}},}\ }\href {\doibase 10.1103/PhysRevB.94.134107} {\bibfield
  {journal} {\bibinfo  {journal} {Physical Review B}\ }\textbf {\bibinfo
  {volume} {94}},\ \bibinfo {pages} {1--12} (\bibinfo {year}
  {2016})}\BibitemShut {NoStop}%
\bibitem [{\citenamefont {Helstrom}(1976)}]{Helstrom1976}%
  \BibitemOpen
  \bibfield  {author} {\bibinfo {author} {\bibfnamefont {C.~W.}\ \bibnamefont
  {Helstrom}},\ }\href@noop {} {\emph {\bibinfo {title} {Quantum Detection and
  Estimation Theory}}}\ (\bibinfo  {publisher} {Academic},\ \bibinfo {address}
  {New York},\ \bibinfo {year} {1976})\BibitemShut {NoStop}%
\bibitem [{\citenamefont {Holevo}(1982)}]{Holevo1982}%
  \BibitemOpen
  \bibfield  {author} {\bibinfo {author} {\bibfnamefont {A.~S.}\ \bibnamefont
  {Holevo}},\ }\href@noop {} {\emph {\bibinfo {title} {Probabilistic and
  Statistical Aspects of Quantum Theory}}}\ (\bibinfo  {publisher} {North
  Holland},\ \bibinfo {address} {Amsterdam},\ \bibinfo {year}
  {1982})\BibitemShut {NoStop}%
\bibitem [{\citenamefont {Bergou}(2015)}]{Bergou2015}%
  \BibitemOpen
  \bibfield  {author} {\bibinfo {author} {\bibfnamefont {Janos~A}\ \bibnamefont
  {Bergou}},\ }\bibfield  {title} {\enquote {\bibinfo {title} {{Quantum state
  discrimination and its applications}},}\ }\href {\doibase
  10.1088/1751-8113/48/8/083001} {\bibfield  {journal} {\bibinfo  {journal}
  {Journal of Physics A: Mathematical and Theoretical}\ }\textbf {\bibinfo
  {volume} {48}},\ \bibinfo {pages} {083001} (\bibinfo {year}
  {2015})}\BibitemShut {NoStop}%
\bibitem [{\citenamefont {Bergou}(2010)}]{Bergou2010}%
  \BibitemOpen
  \bibfield  {author} {\bibinfo {author} {\bibfnamefont {J{\'{a}}nos~A.}\
  \bibnamefont {Bergou}},\ }\bibfield  {title} {\enquote {\bibinfo {title}
  {{Discrimination of quantum states}},}\ }\href {\doibase
  10.1080/09500340903477756} {\bibfield  {journal} {\bibinfo  {journal}
  {Journal of Modern Optics}\ }\textbf {\bibinfo {volume} {57}},\ \bibinfo
  {pages} {160--180} (\bibinfo {year} {2010})}\BibitemShut {NoStop}%
\bibitem [{\citenamefont {Atassi}\ \emph {et~al.}(1998)\citenamefont {Atassi},
  \citenamefont {Chauvin}, \citenamefont {Delaire}, \citenamefont {Delouis},
  \citenamefont {Fanton-Maltey},\ and\ \citenamefont {Nakatani}}]{Atassi1998}%
  \BibitemOpen
  \bibfield  {author} {\bibinfo {author} {\bibfnamefont {Yomen}\ \bibnamefont
  {Atassi}}, \bibinfo {author} {\bibfnamefont {Jerome}\ \bibnamefont
  {Chauvin}}, \bibinfo {author} {\bibfnamefont {J.~A.}\ \bibnamefont
  {Delaire}}, \bibinfo {author} {\bibfnamefont {J.-F.}\ \bibnamefont
  {Delouis}}, \bibinfo {author} {\bibfnamefont {Isabelle}\ \bibnamefont
  {Fanton-Maltey}}, \ and\ \bibinfo {author} {\bibfnamefont {K.}~\bibnamefont
  {Nakatani}},\ }\bibfield  {title} {\enquote {\bibinfo {title} {Photoinduced
  manipulations of photochromes in polymers: Anisotropy, modulation of the nlo
  properties and creation of surface gratings},}\ }\href@noop {} {\bibfield
  {journal} {\bibinfo  {journal} {Pure and Applied Chemistry}\ }\textbf
  {\bibinfo {volume} {70}},\ \bibinfo {pages} {2157} (\bibinfo {year}
  {1998})}\BibitemShut {NoStop}%
\bibitem [{\citenamefont {Barrett}\ \emph {et~al.}(2007)\citenamefont
  {Barrett}, \citenamefont {ichi Mamiya}, \citenamefont {Yager},\ and\
  \citenamefont {Ikeda}}]{Barrett2007}%
  \BibitemOpen
  \bibfield  {author} {\bibinfo {author} {\bibfnamefont {Christopher~J.}\
  \bibnamefont {Barrett}}, \bibinfo {author} {\bibfnamefont {Jun}\ \bibnamefont
  {ichi Mamiya}}, \bibinfo {author} {\bibfnamefont {Kevin~G.}\ \bibnamefont
  {Yager}}, \ and\ \bibinfo {author} {\bibfnamefont {Tomiki}\ \bibnamefont
  {Ikeda}},\ }\bibfield  {title} {\enquote {\bibinfo {title} {Photo-mechanical
  effects in azobenzene-containing soft materials},}\ }\href@noop {} {\bibfield
   {journal} {\bibinfo  {journal} {Soft Matter}\ }\textbf {\bibinfo {volume}
  {3}},\ \bibinfo {pages} {1249--1261} (\bibinfo {year} {2007})}\BibitemShut
  {NoStop}%
\bibitem [{\citenamefont {Kim}\ \emph {et~al.}(2012)\citenamefont {Kim},
  \citenamefont {Safron}, \citenamefont {Huang}, \citenamefont {Arnold},\ and\
  \citenamefont {Gopalan}}]{Kim2012}%
  \BibitemOpen
  \bibfield  {author} {\bibinfo {author} {\bibfnamefont {Myungwoong}\
  \bibnamefont {Kim}}, \bibinfo {author} {\bibfnamefont {Nathaniel~S.}\
  \bibnamefont {Safron}}, \bibinfo {author} {\bibfnamefont {Changshui}\
  \bibnamefont {Huang}}, \bibinfo {author} {\bibfnamefont {Michael~S.}\
  \bibnamefont {Arnold}}, \ and\ \bibinfo {author} {\bibfnamefont {Padma}\
  \bibnamefont {Gopalan}},\ }\bibfield  {title} {\enquote {\bibinfo {title}
  {Light-driven reversible modulation of doping in graphene},}\ }\href@noop {}
  {\bibfield  {journal} {\bibinfo  {journal} {Nano Lett.}\ }\textbf {\bibinfo
  {volume} {12}},\ \bibinfo {pages} {182--187} (\bibinfo {year}
  {2012})}\BibitemShut {NoStop}%
\bibitem [{\citenamefont {Shashikala}\ \emph {et~al.}(2012)\citenamefont
  {Shashikala}, \citenamefont {Nicolas},\ and\ \citenamefont
  {Wang}}]{Shashikala2012}%
  \BibitemOpen
  \bibfield  {author} {\bibinfo {author} {\bibfnamefont {H.~B.~Mihiri}\
  \bibnamefont {Shashikala}}, \bibinfo {author} {\bibfnamefont {Chantel~I.}\
  \bibnamefont {Nicolas}}, \ and\ \bibinfo {author} {\bibfnamefont {Xiao-Qian}\
  \bibnamefont {Wang}},\ }\bibfield  {title} {\enquote {\bibinfo {title}
  {Tunable doping in graphene by light-switchable molecules},}\ }\href@noop {}
  {\bibfield  {journal} {\bibinfo  {journal} {J Phys Chem C Nanomater
  Interfaces}\ }\textbf {\bibinfo {volume} {116}},\ \bibinfo {pages}
  {26102--26105} (\bibinfo {year} {2012})}\BibitemShut {NoStop}%
\bibitem [{\citenamefont {L\'eonard}\ \emph {et~al.}(2017)\citenamefont
  {L\'eonard}, \citenamefont {Spataru}, \citenamefont {Goldflam}, \citenamefont
  {Peters},\ and\ \citenamefont {Beechem}}]{Leonard2017}%
  \BibitemOpen
  \bibfield  {author} {\bibinfo {author} {\bibfnamefont {Fran\c{c}ois}\
  \bibnamefont {L\'eonard}}, \bibinfo {author} {\bibfnamefont {Catalin~D.}\
  \bibnamefont {Spataru}}, \bibinfo {author} {\bibfnamefont {Michael}\
  \bibnamefont {Goldflam}}, \bibinfo {author} {\bibfnamefont {David~W.}\
  \bibnamefont {Peters}}, \ and\ \bibinfo {author} {\bibfnamefont {Thomas~E.}\
  \bibnamefont {Beechem}},\ }\bibfield  {title} {\enquote {\bibinfo {title}
  {Dynamic wavelength-tunable photodetector using subwavelength graphene
  field-effect transistors},}\ }\href@noop {} {\bibfield  {journal} {\bibinfo
  {journal} {Scientific Reports}\ }\textbf {\bibinfo {volume} {7}},\ \bibinfo
  {pages} {45873} (\bibinfo {year} {2017})}\BibitemShut {NoStop}%
\bibitem [{\citenamefont {Loubser}\ and\ \citenamefont {van
  Wyk}(1977)}]{Loubser1977}%
  \BibitemOpen
  \bibfield  {author} {\bibinfo {author} {\bibfnamefont {J.~H.~N.}\
  \bibnamefont {Loubser}}\ and\ \bibinfo {author} {\bibfnamefont {J.~A.}\
  \bibnamefont {van Wyk}},\ }\bibfield  {title} {\enquote {\bibinfo {title}
  {Diamond research},}\ }\href@noop {} {\bibfield  {journal} {\bibinfo
  {journal} {Industial Diamond Information Bureau}\ }\textbf {\bibinfo {volume}
  {11}},\ \bibinfo {pages} {4} (\bibinfo {year} {1977})}\BibitemShut {NoStop}%
\bibitem [{\citenamefont {Loubser}\ and\ \citenamefont {van
  Wyk}(1978)}]{Loubser1978}%
  \BibitemOpen
  \bibfield  {author} {\bibinfo {author} {\bibfnamefont {J.~H.~N.}\
  \bibnamefont {Loubser}}\ and\ \bibinfo {author} {\bibfnamefont {J.~A.}\
  \bibnamefont {van Wyk}},\ }\bibfield  {title} {\enquote {\bibinfo {title}
  {Electron spin resonance in the study of diamond},}\ }\href@noop {}
  {\bibfield  {journal} {\bibinfo  {journal} {Rep. Prog. Phys.}\ }\textbf
  {\bibinfo {volume} {41}},\ \bibinfo {pages} {1201} (\bibinfo {year}
  {1978})}\BibitemShut {NoStop}%
\bibitem [{\citenamefont {{Van Oort}}\ and\ \citenamefont
  {Glasbeek}(1990)}]{vanOort1990}%
  \BibitemOpen
  \bibfield  {author} {\bibinfo {author} {\bibfnamefont {Eric}\ \bibnamefont
  {{Van Oort}}}\ and\ \bibinfo {author} {\bibfnamefont {Max}\ \bibnamefont
  {Glasbeek}},\ }\bibfield  {title} {\enquote {\bibinfo {title}
  {{Electric-field-induced modulation of spin echoes of N-V centers in
  diamond}},}\ }\href {\doibase 10.1016/0009-2614(90)85665-Y} {\bibfield
  {journal} {\bibinfo  {journal} {Chemical Physics Letters}\ }\textbf {\bibinfo
  {volume} {168}},\ \bibinfo {pages} {529--532} (\bibinfo {year}
  {1990})}\BibitemShut {NoStop}%
\bibitem [{\citenamefont {Balasubramanian}\ \emph {et~al.}(2009)\citenamefont
  {Balasubramanian}, \citenamefont {Neumann}, \citenamefont {Twitchen},
  \citenamefont {Markham}, \citenamefont {Kolesov}, \citenamefont {Mizuochi},
  \citenamefont {Isoya}, \citenamefont {Achard}, \citenamefont {Beck},
  \citenamefont {Tissler}, \citenamefont {Jacques}, \citenamefont {Hemmer},
  \citenamefont {Jelezko},\ and\ \citenamefont
  {Wrachtrup}}]{Balasubramanian2009}%
  \BibitemOpen
  \bibfield  {author} {\bibinfo {author} {\bibfnamefont {Gopalakrishnan}\
  \bibnamefont {Balasubramanian}}, \bibinfo {author} {\bibfnamefont {Philipp}\
  \bibnamefont {Neumann}}, \bibinfo {author} {\bibfnamefont {Daniel}\
  \bibnamefont {Twitchen}}, \bibinfo {author} {\bibfnamefont {Matthew}\
  \bibnamefont {Markham}}, \bibinfo {author} {\bibfnamefont {Roman}\
  \bibnamefont {Kolesov}}, \bibinfo {author} {\bibfnamefont {Norikazu}\
  \bibnamefont {Mizuochi}}, \bibinfo {author} {\bibfnamefont {Junichi}\
  \bibnamefont {Isoya}}, \bibinfo {author} {\bibfnamefont {Jocelyn}\
  \bibnamefont {Achard}}, \bibinfo {author} {\bibfnamefont {Johannes}\
  \bibnamefont {Beck}}, \bibinfo {author} {\bibfnamefont {Julia}\ \bibnamefont
  {Tissler}}, \bibinfo {author} {\bibfnamefont {Vincent}\ \bibnamefont
  {Jacques}}, \bibinfo {author} {\bibfnamefont {Philip~R}\ \bibnamefont
  {Hemmer}}, \bibinfo {author} {\bibfnamefont {Fedor}\ \bibnamefont {Jelezko}},
  \ and\ \bibinfo {author} {\bibfnamefont {Joerg}\ \bibnamefont {Wrachtrup}},\
  }\bibfield  {title} {\enquote {\bibinfo {title} {Ultralong spin coherence
  time in isotopically engineered diamond},}\ }\href {\doibase
  10.1038/NMAT2420} {\bibfield  {journal} {\bibinfo  {journal} {Nature
  Materials}\ }\textbf {\bibinfo {volume} {8}},\ \bibinfo {pages} {383--387}
  (\bibinfo {year} {2009})}\BibitemShut {NoStop}%
\bibitem [{\citenamefont {Rosskopf}\ \emph {et~al.}(2014)\citenamefont
  {Rosskopf}, \citenamefont {Dussaux}, \citenamefont {Ohashi}, \citenamefont
  {Loretz}, \citenamefont {Schirhagl}, \citenamefont {Watanabe}, \citenamefont
  {Shikata}, \citenamefont {Itoh},\ and\ \citenamefont {Degen}}]{Rosskopf2014}%
  \BibitemOpen
  \bibfield  {author} {\bibinfo {author} {\bibfnamefont {T.}~\bibnamefont
  {Rosskopf}}, \bibinfo {author} {\bibfnamefont {A.}~\bibnamefont {Dussaux}},
  \bibinfo {author} {\bibfnamefont {K.}~\bibnamefont {Ohashi}}, \bibinfo
  {author} {\bibfnamefont {M.}~\bibnamefont {Loretz}}, \bibinfo {author}
  {\bibfnamefont {R.}~\bibnamefont {Schirhagl}}, \bibinfo {author}
  {\bibfnamefont {H.}~\bibnamefont {Watanabe}}, \bibinfo {author}
  {\bibfnamefont {S.}~\bibnamefont {Shikata}}, \bibinfo {author} {\bibfnamefont
  {K.~M.}\ \bibnamefont {Itoh}}, \ and\ \bibinfo {author} {\bibfnamefont
  {C.~L.}\ \bibnamefont {Degen}},\ }\bibfield  {title} {\enquote {\bibinfo
  {title} {{Investigation of surface magnetic noise by shallow spins in
  diamond}},}\ }\href {\doibase 10.1103/PhysRevLett.112.147602} {\bibfield
  {journal} {\bibinfo  {journal} {Physical Review Letters}\ }\textbf {\bibinfo
  {volume} {112}},\ \bibinfo {pages} {1--5} (\bibinfo {year}
  {2014})}\BibitemShut {NoStop}%
\bibitem [{\citenamefont {Szankowski}\ \emph {et~al.}(2013)\citenamefont
  {Szankowski}, \citenamefont {Trippenbach},\ and\ \citenamefont
  {Band}}]{Szankowski2013}%
  \BibitemOpen
  \bibfield  {author} {\bibinfo {author} {\bibfnamefont {Piotr}\ \bibnamefont
  {Szankowski}}, \bibinfo {author} {\bibfnamefont {M.}~\bibnamefont
  {Trippenbach}}, \ and\ \bibinfo {author} {\bibfnamefont {Y.~B.}\ \bibnamefont
  {Band}},\ }\bibfield  {title} {\enquote {\bibinfo {title} {Spin decoherence
  due to fluctuating fields},}\ }\href {\doibase 10.1103/PhysRevE.87.052112}
  {\bibfield  {journal} {\bibinfo  {journal} {Phys. Rev. E}\ }\textbf {\bibinfo
  {volume} {87}},\ \bibinfo {pages} {052112} (\bibinfo {year}
  {2013})}\BibitemShut {NoStop}%
\bibitem [{\citenamefont {Collins}\ \emph {et~al.}(1983)\citenamefont
  {Collins}, \citenamefont {Thomaz},\ and\ \citenamefont
  {Jorge}}]{Collins1983}%
  \BibitemOpen
  \bibfield  {author} {\bibinfo {author} {\bibfnamefont {A.~T.}\ \bibnamefont
  {Collins}}, \bibinfo {author} {\bibfnamefont {M.~F.}\ \bibnamefont {Thomaz}},
  \ and\ \bibinfo {author} {\bibfnamefont {M.~I.~B.}\ \bibnamefont {Jorge}},\
  }\bibfield  {title} {\enquote {\bibinfo {title} {Luminescence decay time of
  the 1.945 \hbox{eV} centre in type \hbox{Ib} diamond},}\ }\href@noop {}
  {\bibfield  {journal} {\bibinfo  {journal} {Journal of Physics C}\ }\textbf
  {\bibinfo {volume} {16}},\ \bibinfo {pages} {2177} (\bibinfo {year}
  {1983})}\BibitemShut {NoStop}%
\bibitem [{\citenamefont {Hanzawa}\ \emph {et~al.}(1997)\citenamefont
  {Hanzawa}, \citenamefont {Nisida},\ and\ \citenamefont {Kato}}]{Hanzawa1997}%
  \BibitemOpen
  \bibfield  {author} {\bibinfo {author} {\bibfnamefont {J.}~\bibnamefont
  {Hanzawa}}, \bibinfo {author} {\bibfnamefont {Y.}~\bibnamefont {Nisida}}, \
  and\ \bibinfo {author} {\bibfnamefont {T.}~\bibnamefont {Kato}},\ }\bibfield
  {title} {\enquote {\bibinfo {title} {Measurement of decay time for the
  \hbox{NV} centre in \hbox{Ib} diamond with a picosecond laser pulse},}\
  }\href@noop {} {\bibfield  {journal} {\bibinfo  {journal} {Diamond and
  Related Materials}\ }\textbf {\bibinfo {volume} {6}},\ \bibinfo {pages}
  {1595} (\bibinfo {year} {1997})}\BibitemShut {NoStop}%
\bibitem [{\citenamefont {Happ}\ and\ \citenamefont
  {Freyberger}(2008)}]{Happ2008}%
  \BibitemOpen
  \bibfield  {author} {\bibinfo {author} {\bibfnamefont {Christof~J.}\
  \bibnamefont {Happ}}\ and\ \bibinfo {author} {\bibfnamefont {Matthias}\
  \bibnamefont {Freyberger}},\ }\bibfield  {title} {\enquote {\bibinfo {title}
  {{Adaptive estimation of qubits by symmetry measurements}},}\ }\href
  {\doibase 10.1103/PhysRevA.78.064303} {\bibfield  {journal} {\bibinfo
  {journal} {Physical Review A - Atomic, Molecular, and Optical Physics}\
  }\textbf {\bibinfo {volume} {78}},\ \bibinfo {pages} {1--4} (\bibinfo {year}
  {2008})},\ \Eprint {http://arxiv.org/abs/0811.2360} {0811.2360} \BibitemShut
  {NoStop}%
\bibitem [{\citenamefont {Higgins}\ \emph {et~al.}(2011)\citenamefont
  {Higgins}, \citenamefont {Doherty}, \citenamefont {Bartlett}, \citenamefont
  {Pryde},\ and\ \citenamefont {Wiseman}}]{Higgins2011}%
  \BibitemOpen
  \bibfield  {author} {\bibinfo {author} {\bibfnamefont {B.~L.}\ \bibnamefont
  {Higgins}}, \bibinfo {author} {\bibfnamefont {A.~C.}\ \bibnamefont
  {Doherty}}, \bibinfo {author} {\bibfnamefont {S.~D.}\ \bibnamefont
  {Bartlett}}, \bibinfo {author} {\bibfnamefont {G.~J.}\ \bibnamefont {Pryde}},
  \ and\ \bibinfo {author} {\bibfnamefont {H.~M.}\ \bibnamefont {Wiseman}},\
  }\bibfield  {title} {\enquote {\bibinfo {title} {{Multiple-copy state
  discrimination: Thinking globally, acting locally}},}\ }\href {\doibase
  10.1103/PhysRevA.83.052314} {\bibfield  {journal} {\bibinfo  {journal}
  {Physical Review A - Atomic, Molecular, and Optical Physics}\ }\textbf
  {\bibinfo {volume} {83}},\ \bibinfo {pages} {1--10} (\bibinfo {year}
  {2011})}\BibitemShut {NoStop}%
\bibitem [{\citenamefont {Wang}\ \emph {et~al.}(2017)\citenamefont {Wang},
  \citenamefont {Li},\ and\ \citenamefont {Zou}}]{Wang2017}%
  \BibitemOpen
  \bibfield  {author} {\bibinfo {author} {\bibfnamefont {Yuan-Mei}\
  \bibnamefont {Wang}}, \bibinfo {author} {\bibfnamefont {Jun-Gang}\
  \bibnamefont {Li}}, \ and\ \bibinfo {author} {\bibfnamefont {Jian}\
  \bibnamefont {Zou}},\ }\bibfield  {title} {\enquote {\bibinfo {title}
  {{Detecting the presence of a magnetic field under Gaussian and non-Gaussian
  noise by adaptive measurement}},}\ }\href {\doibase
  10.1016/j.physleta.2017.04.003} {\bibfield  {journal} {\bibinfo  {journal}
  {Physics Letters A}\ }\textbf {\bibinfo {volume} {381}},\ \bibinfo {pages}
  {1866--1873} (\bibinfo {year} {2017})}\BibitemShut {NoStop}%
\bibitem [{\citenamefont {Wang}(1993)}]{Wang1993}%
  \BibitemOpen
  \bibfield  {author} {\bibinfo {author} {\bibfnamefont {Y.~H.}\ \bibnamefont
  {Wang}},\ }\bibfield  {title} {\enquote {\bibinfo {title} {{On the Number of
  Successes in Independent Trials}},}\ }\href@noop {} {\bibfield  {journal}
  {\bibinfo  {journal} {Statistica Sinica}\ }\textbf {\bibinfo {volume} {3}},\
  \bibinfo {pages} {295--312} (\bibinfo {year} {1993})}\BibitemShut {NoStop}%
\end{thebibliography}

%

\end{document}